\documentclass{aa} 

\usepackage{longtable,lscape}
\usepackage[figuresright]{rotating}
\usepackage{epsfig}
\usepackage{graphicx}
\usepackage{natbib}
\bibpunct{(}{)}{;}{a}{}{,} 
\usepackage{txfonts}
\def\sun{\hbox{$_{\odot}$}}

\begin{document}

\title{An updated catalog of OH-maser-emitting planetary nebulae}

\author{L. Uscanga \inst{1,2}
        \and J.F. G\'omez \inst{2}
        \and O. Su\'arez \inst{3}
        \and L.F. Miranda \inst{4,5}\thanks{present address: Universidade de Vigo}
        }

\offprints{L. Uscanga}

\institute{Observatorio Astron\'omico Nacional (IGN), 
Alfonso XII No. 3, E-28014 Madrid, Spain  email: l.uscanga@oan.es
\and 
Instituto de Astrof\'{\i}sica de Andaluc\'{\i}a, CSIC, Apartado
  3004, E-18080 Granada, Spain 
\and
Laboratoire Lagrange, UMR7293, Universit\'e de Nice Sophia-Antipolis, CNRS, Observatoire de la C\^ote d'Azur, 
06304 Nice Cedex 4, France
\and
Consejo Superior de Investigaciones Cient\'ificas, C/ Serrano 117, E-28006, Madrid, Spain
\and 
Departamento de Fisica Aplicada, Facultade de Ciencias, Universidade de Vigo, E-36310 Vigo, Spain 
}


\abstract
   {} 
{We studied the characteristics of planetary nebulae (PNe) that show
  both OH maser and radio   
continuum emission (hereafter OHPNe). These have been proposed to be very young
PNe, and therefore, they could be key objects for understanding the
formation and evolution of PNe.}
{We consulted the literature searching for interferometric observations
  of radio continuum and OH masers toward evolved stars, including the
  information from several surveys. We also processed
  radio continuum and OH maser observations toward PNe
  in the  Very Large Array data archive.
  The high positional accuracy provided by interferometric observations allow us to confirm or reject the association between OH maser and radio continuum emission. }
    {We found a total of six PNe that present both
     OH maser and radio continuum emissions, as confirmed with radio 
     interferometric observations. These are bona fide OHPNe. 
     The confirmed OHPNe present a bipolar morphology in resolved images of their 
     ionized emission at different wavelengths, 
     suggesting that the OH maser emission in PNe
    is related to nonspherical mass-loss phenomena. 
     The OH maser spectra in PNe present a clear asymmetry, tending to show
    blueshifted emission with respect to the systemic velocity. 
    Their infrared colors suggest that most of these objects are very young PNe.
    OHPNe do not form a homogeneous group, 
    and seem to represent a variety of different evolutionary stages. We suggest that OH masers pumped in the 
    AGB phase may disappear during the post-AGB phase, but reappear once the source becomes a PN 
    and its radio continuum emission is amplified by the OH molecules. Therefore, OH maser emission 
    could last significantly longer than the previously assumed 1000 yr after the end of the AGB phase. 
    This maser lifetime may be longer in PNe with more massive central stars, 
    which ionize a larger amount of gas in the envelope. 
   } 
    {}

\keywords{masers -- stars: AGB and post-AGB -- ISM: planetary nebulae: general}

\maketitle
%
%
\section{Introduction}

Planetary nebulae (PNe) are a phase in the evolution of low and intermediate
mass stars ($\le8M\sun$), after they have passed through both the asymptotic giant
branch (AGB) phase and a brief, 10$^2$ to 10$^4$ yr, post-AGB phase \citep{blocker95}.
The effective temperature of the central star increases rapidly during this last
phase, until at $\sim$30,000 K its radiation begins to ionize the circumstellar
shell, whereupon it becomes a PN \citep{kwok93}. The ionized gas of PNe is detected
at radio wavelengths from its free-free radiation. Radio continuum observations
thus provide an important diagnostic for PNe per se, as well as information
about their physical structure and properties (e.g., electron density, mass).

The circumstellar envelopes of oxygen-rich evolved stars provide
optimal conditions for pumping masers of different species. Maser emission of 
SiO, H$_2$O, and OH is detected from AGB 
envelopes, and it tends to be stratified, with SiO masers located close to the central
star ($\sim$10 AU), H$_2$O masers located in the inner part of the
envelope (between 10 and 100 AU) and OH masers at distances $> 10^3$
AU from the central star \citep{reidmoran81}. In the particular case
of OH maser emission, the spectra usually show the double-peaked profiles
characteristic of spherical envelopes expanding at velocities of
$\sim$10$-$30 km~s$^{-1}$ \citep{reid77,bowers89}.  

OH masers are expected to disappear $\simeq$1000 yr after the AGB
mass-loss stops \citep{lewis89,gomezy90}. Consequently, in some early
PNe, the inner part of the 
envelope could already be ionized, while the outer regions are still
neutral. These objects could exhibit OH maser emission from the
neutral region, as well as radio continuum emission from the inner
ionized envelope.  
Therefore, they would
show characteristics of both standard OH/IR stars (strong infrared
emission together with OH maser emission) and PNe (radio continuum
emission at centimeter wavelengths). These peculiar sources
are called OHPNe \citep{zijlstra89}. 

This rare type of source probably represents an evolutionary phase
immediately before the formation of a full-blown PN
\citep{zijlstra89}.  
Therefore these sources are potentially key objects for studying the early evolution of PNe. 
 \citet{garciahernandez07} propose that strongly obscured OHPNe
 represent the population of high-mass precursors of PNe in our
 Galaxy. Given their potential interest in studying the early phases of
 PN evolution, it is important to have an updated and reliable catalog
 of OHPNe in order to study their characteristics and the possible
 relationship with other maser-emitting evolved objects. 

The seminal work by \citet{zijlstra89} cataloged 12 possible OHPNe. These were 
identified using three tools: (1) radio continuum observations of evolved stars 
with reported OH emission, (2) OH observations of possible young PNe, and (3) 
correlations between the OH and PN catalogs.
However, as these authors point out, in several of those 12 identified candidates, the association
between OH maser and radio continuum could not be confirmed, owing to
the low positional accuracy of the data. Only in four of these objects had
the positions of OH 
maser emission been measured with interferometric observations at that time. In
the rest of the sources, the OH maser data were obtained in
single-dish observations with beams of several arcminutes, so it is
hard to ascertain whether the maser emission arises from the same
source that shows radio continuum emission. In fact, in two of the sources
cataloged by \citet{zijlstra89}, new interferometric observations by
\citet{gomez08} 
show that the reported continuum and OH maser emission were not
associated with each other, therefore their classification as OHPNe should be
revised. 

After the publication by \citet{zijlstra89}, new interferometric
observations of radio continuum and OH masers toward evolved stars
have been carried out, including large-scale surveys, e.g., the NRAO
VLA Sky Survey (NVSS) \citep{condon98a} and the ATCA/VLA OH 1612 MHz survey
\citep{sevenster97a,sevenster97b,sevenster01}. Therefore, it is now
possible to   update the catalog of OHPNe, confirming 
the possible candidates with the current available radio-continuum and
OH maser data of interferometric observations, which provide
sufficient positional accuracy to associate these emissions. 

Another important aspect to consider is the actual nature of these sources. 
We note that the observational definition of OHPNe as sources with
both radio continuum and OH maser emissions \citep{zijlstra89} may
include sources that are not really PNe, so labeling them collectively 
as OHPNe may 
be misleading.  There are evolved stars that show radio continuum 
emission and OH maser emission but are not PNe, such as symbiotic systems
 \citep{kafatos89,aaquist90,ivison94}, in which the envelope of a red
 giant is ionized by a hot companion (e.g., a white dwarf) or post-AGB
 stars \citep{bains09} still too cold to ionize their envelopes but that
may show radio continuum emission
 arising from shocks. If we want to study OHPNe in the context of the
 early evolution of PNe, it is important to distinguish between
 different types of 
 OH and radio continuum emitters. In this paper, we reserve the
 term OHPNe for bona fide PNe only (objects in which photoionization
 from the central star has started), separating them from other evolved
 objects with OH maser and radio continuum emission.

In this work we present an updated list of OHPNe, with a study on
their nature and characteristics.
This paper is organized as follows.  
In Sect. \ref{description} we describe the analysis of radio interferometric archival data of both radio continuum and OH maser emission. 
In Sect. \ref{results} we describe the results obtained, including
a description of confirmed sources, possible candidates, misclassified 
 sources as OHPNe, and related objects.
In Sect. \ref{characteristics} we summarize the general
characteristics of the OHPNe sources. 
In Sect. \ref{discussion} we present a brief discussion,
and finally, in Sect. \ref{summary} we give our main conclusions.

\section{Description of the analysis of radio interferometric data}
\label{description}

\subsection{Archival data of cataloged OHPNe}

We retrieved  and processed radio continuum and OH maser 
data from the 12
candidate OHPN reported by \citet{zijlstra89} that were available in
  the Very Large Array archive of the  National Radio Astronomy Observatory
(NRAO) \footnote{The National Radio Astronomy Observatory is a
  facility of the National Science Foundation operated under
  cooperative agreement by Associated Universities, Inc.}. Our goal
  was to confirm the association between maser and radio continuum
  emission in those candidate sources. A list of the archival data used is given in
  Table \ref{archivaldata}.

VLA continuum and OH line data were calibrated and processed using the
Astronomical Image Processing System (AIPS) package of NRAO, following
 standard procedures. Some data presented here have been previously
published
\citep{pottasch87,pottasch88,zijlstra89,ratag90}. However, our new data
reduction includes the use of the robust scheme of visibility
weighting \citep{briggs95}, not available at the time of the original
publications. We used a ROBUST parameter equals 0, as implemented
in task IMAGR of AIPS, since it optimizes the trade-off between
angular resolution and point-source sensitivity. Positions of
continuum and OH emission were obtained by fitting elliptical Gaussians
with task JMFIT. Unless stated otherwise, the positions given in this
paper are those obtained in our reanalysis of data. 

\subsection{Literature search for new candidates}

To find new OHPN candidate objects, 
additional to the ones reported by \citet{zijlstra89}, we consulted the
literature searching for interferometric data of both radio continuum
and OH masers in PNe, identifying coincidences between their
positions. 

Positional information on the radio continuum emission was retrieved
from published catalogs of optically identified PNe with radio continuum
counterparts detected in large-scale interferometric surveys. 
In particular, we used the list of NVSS radio continuum 
detections \citep{condon98} coincident with sources in the Strasbourg-ESO Catalog of Galactic
Planetary Nebulae \citep{acker92}, and in subsequent optical surveys
\citep{luo05}. Furthermore, we used the PNe from the 
 the new Macquarie/AAO/Strasbourg H$\alpha$
(MASH) catalog detected at centimeter wavelengths in the NVSS and the Molonglo
Galactic Plane Survey (MGPS-2), reported by
\citet{bojicic11}. Other PNe with
radio continuum observations reported elsewhere
\citep{aaquist90,vandsteene93,vandsteene95,vandsteene01} were also
included in our list.

For positional information on OH maser emission, we used the ATCA/VLA OH
1612 MHz survey \citep{sevenster97a,sevenster97b,sevenster01}, and 
publications reporting interferometric OH observations 
on specific sources, compiled in the 
database of circumstellar OH masers of \citet{engels10}, which is a very 
complete compilation of OH maser data in evolved stars.
With the positions of radio-continuum emission associated with PNe and
the accurate positions of OH masers provided by the interferometric
observations and surveys mentioned above, we  made
cross-identifications to find possible new OHPNe. 

\subsection{Cross-identification of radio continuum and maser emission}
\label{association}

From both our literature search and our own analysis of VLA
archive data, we compared the positions of the OH maser emission with
those of the radio continuum, to ascertain in which sources those emissions
are associated. In the case of resolved radio continuum
emission, we consider them associated if the maser position is
projected within the
limits of the radio continuum. For unresolved emission, we consider
that the radio continuum and maser emission are associated when their
positions are compatible within their relative positional errors.

For nonsimultaneous radio continuum and OH maser observations,
the relative errors are imposed by the absolute astrometric error of
each observation. These errors are difficult to evaluate in an
interferometric observation (especially when using positions from the
literature),  since they depend on a number of factors,
such as the phase errors, the intrinsic positional error of the phase
calibrator, the distance from the target source to that calibrator, 
or the accuracy in the measurements of baseline
length. Typically, these errors could be around one tenth of the
synthesized beam \citep{baudry01}. In this paper we have
conservatively assumed positional errors of one fifth of the synthesized
beam, and are typically $<3''$. In addition to these errors,
intrinsic to interferometric observations, there is a positional
uncertainty due to the noise in the images \citep{reid88}, which is $\sigma_n \simeq
\frac{\theta}{2 snr}$, where $\sigma_n$ is the 1-sigma uncertainty due
to noise,
$\theta$ the synthesized beam, and $snr$ the signal-to-noise
ratio of the source. In our case, this error due to noise is always much smaller
than the former component, so it is negligible in comparison.

However, we note that in some data sets used in this paper, there are
simultaneous observations of OH masers and radio continuum at 18~cm. 
In those cases, the relative uncertainties are determined by the
noise error alone ($\sigma_n$), since the other components of absolute errors are
the same for maser and continuum images. Therefore, for these simultaneous
observations, the relative positions can be measured with much
higher accuracy.

\section{Results}
\label{results}

The result of our reduction of archival data was 
the confirmation of five objects as OHPNe, and the rejection of 
four of them, from the list of \citet{zijlstra89}. The additional literature search 
yielded 
a confirmed OHPN (IRAS 19255+2123) and a possible one (IRAS 17168$-$3736).

\subsection{Confirmed OHPNe}
We consider as true OHPNe only those objects identified in the literature as
PNe in which both the OH
maser and the radio continuum emissions have been detected, and these
emissions are associated with the criteria mentioned in Sect.
\ref{association}. Evolved objects with associated OH and radio
continuum, but not found to be PNe, are not considered in this category
(presented as related objects in Sect. \ref{related}).
We have confirmed a total of six OHPNe, which are listed in Table 
\ref{confirmations}.
In the following, we give a brief description of each source.  

\subsubsection{IRAS 17103$-$3702 (NGC 6302)}
NGC 6302 is a well-known multipolar PN. The 
H$\alpha$ and [NII] images show a complex morphology, with 
ionized lobes extended roughly in the east-west
direction and a dark lane covering the waist of the nebula
\citep[see][and references therein]{meaburn05}. 
This source was listed by \citet{zijlstra89} as OHPN~1.
The first simultaneous VLA observations of continuum emission and OH
maser line at 1612 MHz in this PN were reported by
\citet{payne88}. These observations clearly show the association of
the OH maser emission with the continuum emission \citep[see Figure 4
of][]{payne88}, with the OH masers being projected against the
resolved radio
continuum. Since the continuum and maser observations were
simultaneous and at close frequencies, this spatial association is
highly reliable.
The radio continuum map presents a bipolar morphology better
shown at 6~cm \citep{gomez89}, although extending along the northeast-southwest  direction,
different from the one in the optical images.
The OH maser spectrum is dominated by a single peak at $V_{\rm LSR} \simeq 
-41$
km~s$^{-1}$ \citep{payne88},
blueshifted with respect to the velocity of the optical nebula 
\citep[$-$31.4 km~s$^{-1}$,][]{schneider83}. 
The spectrum could be the superposition of several velocity components, 
giving rise to a broad profile, with emission spanning $\ga 20$ km~s$^{-1}$. 
Thereafter, \citet{sevenster97b} also detected this source in the
ATCA/VLA OH 1612 MHz survey toward the galactic disk region, and
\citet{condon98} detected the radio continuum emission at 20~cm in the
NVSS. These positions are separated by only $\simeq 2''$ (see Table \ref{confirmations}), which is within their positional errors. 

\subsubsection{IRAS 17347$-$3139}
This young PN is obscured at optical wavelengths, but shows a clear
bipolar morphology in near-infrared images
\citep{degregorio04,sahai07}. The lobes have a total extension of
$\sim$4$''$ and are separated by a dark equatorial lane. Recent mid-infrared images
with VISIR/VLT reveal a
multipolar structure, possibly indicating
that the nebula has been shaped by precessing jets \citep{lagadec11}.   

This source was listed by \citet{zijlstra89} as OHPN~5. 
The 6~cm continuum of this PN was first
reported by \citet{ratag90} and OH maser emission by \citet{zijlstra89}.  
A detailed study of the radio continuum emission from this source 
has been carried out by \citet{gomez05}, who interpret it as arising from an expanding ionized nebula, with a  dynamical age of $\sim$120 yr. 
Further high-angular resolution observations of
continuum (at 0.7, 1.3, 3.6, and 18~cm) and OH maser emission at 1612
MHz with the VLA were carried out by 
\citet{tafoya09}. The radio continuum emission has a bipolar morphology of $\sim$1\farcs5 in size, 
coinciding with the lobes observed at near-infrared wavelengths.
The OH maser emission shows a narrow single  peak at $V_{\rm
  LSR} = -70$
km~s$^{-1}$, which is spatially projected within the limits
of the radio continuum. This confirms their association, especially 
considering that the OH and continuum data at
18~cm were obtained simultaneously. Some of the additional 
spectral features present in the single-dish OH spectra \citep{zijlstra89} 
may actually arise from other sources in its neighborhood \citep{tafoya09}. 
The OH feature is blueshifted with respect to the systemic velocity, 
$\simeq -55$ km~s$^{-1}$, derived from interferometric CO observations
(D. Tafoya, private communication).

We also note that water maser emission was observed within
$\sim$0\farcs 25 from the center of this OHPN, its association being
firmly established through simultaneous interferometric observations
of water masers and radio continuum at 1.3~cm
\citep{degregorio04}. The spatial distribution and radial velocities
of the water masers are suggestive of a rotating and expanding maser
ring that traces the innermost regions of a torus formed at the end of
the AGB phase.

\subsubsection{JaSt 23} 	
\citet{vandsteene01} detected radio continuum emission at 3 and 6~cm toward
this source  using ATCA (Table \ref{confirmations}).  The
position of this continuum source is offset $\simeq 32''$ from IRAS
17371$-$2747, while the positional error of the IRAS source is
$26''$. Therefore, it is possible that this continuum source is not associated with 
IRAS 17371$-$2747, as previously assumed by \citet{zijlstra89}, who labeled
it as OHPN~6. The cataloged IRAS source may in fact be the superposition of two different sources 
that are
unresolved by IRAS (angular resolution $\simeq 1'-4'$), 
but distinct in WISE images: JaSt 23 (whose counterpart is likely to be 
WISE J174023.06-274911.7) and
another reddened object (WISE J174018.10-274849.3), separated by $\simeq 70''$. 
The position of the radio continuum coincides with that
of the PN JaSt 23 \citep{jacoby04}. 
We also note that the continuum sources
reported by \citet{vandsteene01} and \citet{zijlstra89} are separated by
 more than $1'$, and we could
only confirm the former in our data reprocessing.  
The OH maser emission at 1612 MHz of this source was detected
in the ATCA/VLA survey toward the Galactic bulge region
by \citet{sevenster97a}. The OH maser spectrum shows a single peak at
$V_{\rm
  LSR} =+115.2$ km~s$^{-1}$, with a broad profile (width $\simeq 15$
km~s$^{-1}$).  The separation between the OH maser and continuum 
emissions is only $\sim$0$\farcs 9$ (Table \ref{confirmations}). 
Considering the positional
uncertainties, the OH maser and continuum emissions seem to be
associated.  The systemic velocity is still unknown.

\subsubsection{IRAS 17393$-$2727}
An HST image of IRAS 17393$-$2727 shows a highly collimated bipolar
outflow, as well as a dark lane that completely obscures the central
source \citep{manteiga11}. The infrared spectrum of this source shows
bright [Ne II] nebular emission, strongly suggesting that the ionization of its
central region has already started \citep{garciahernandez07}.

This source was listed by \citet{zijlstra89} as OHPN~9.
The radio continuum emission toward this source was detected at 2 and
6~cm using the VLA by \citet{pottasch87}. The OH maser line at 1612 MHz
was observed 
by  \citet{zijlstra89} also with the VLA. 
We have confirmed the positions of both emissions in our reprocessing
of VLA data archive, as well as OH maser emission at 1665 MHz.
Moreover, \citet{sevenster97a} detected this source in the ATCA/VLA
survey toward the Galactic bulge region. The OH spectrum at 1612 MHz shows a highly asymmetrical double peak, with features located
at $V_{\rm
  LSR} =-122.8$ km~s$^{-1}$ and  $-93.6$ km~s$^{-1}$, and intensity
ratio (blueshifted/redshifted) of $\simeq 70$. The positions of the OH maser and the radio-continuum
emissions are separated $\sim$0\farcs3, and are compatible within the
errors (Table \ref{confirmations}).  

\subsubsection{IRAS 19219$+$0947 (Vy 2$-$2)}
\citet{sahai11} classified this source as an irregular PN, based on an
HST H$\alpha$ image that shows 
extended emission but no obvious structure.
Analysis of the internal kinematics \citep{miranda91} shows 
that Vy 2$-$2 consists of a slow expanding, compact central region ($\simeq$0\farcs35) and two rather wide bipolar features ($\ge$0\farcs6). This is consistent with an
elliptical/bipolar PN and with the structures observed in radio continuum (see below).

This source was listed by \citet{zijlstra89} as OHPN~12.
Vy 2$-$2 is the first PN where OH maser emission was detected 
with single-dish observations \citep{davis79}. 
\citet{seaquist83} detected radio continuum (at 2, 6, and 20~cm) and
OH  (1612 MHz) maser emission with the VLA toward this source. 
Moreover, OH maser emission at 6035 MHz was detected toward this source 
with single-dish observations
\citep{jewell85,desmurs02,desmurs10}. However, 
interferometric observations would be
necessary to confirm its association with Vy 2$-$2. 
The VLA radio continuum map at 6~cm shows a slight elongation in the
northeast-southwest direction \citep{seaquist83,christianto98}. However, at 
1.3~cm \citep{seaquist91} and 2~cm \citep{seaquist83}, 
it has a clear shell-like structure, 
with a diameter of $\simeq 0.4''$.
The radio continuum emission of this source was also  detected in the
NVSS \citep{condon98}. The OH maser spectrum shows a single peak at $V_{\rm
  LSR}\simeq -$62 km~s$^{-1}$ with an asymmetrical profile spanning
$\ga 8$ km~s$^{-1}$, which is blueshifted with respect to the
velocity of the optical nebula \citep[$-53.4$ km~s$^{-1}$,][]{miranda91}. 
The spatio-kinematical distribution of the OH maser emission 
  is also consistent with
  an expanding shell \citep{shepherd90}, with the same center as the one seen in radio continuum.
The distance between the peak of the continuum emission at 20~cm and
the OH masers is only $\sim$0$\farcs$25 (within the observational
errors, Table \ref{confirmations}). Moreover, the peak of the maser emission is projected against
resolved structures traced by the radio continuum at 1.3, 2, and 6~cm
\citep{seaquist83,seaquist91}, clearly indicating that
these emissions are associated.  

\subsubsection{IRAS 19255$+$2123 (K 3$-$35)}

K 3$-$35 is a young PN that shows a bipolar morphology in both optical images
\citep{miranda00} and  
VLA radio continuum maps at 2, 3.6, and 6~cm 
\citep{aaquist89, miranda01}. 
This continuum source was also detected in the NVSS at 20~cm \citep{condon98}. 
The radio continuum emission exhibits a bright core and two
bipolar lobes with an S-shape, which
can be explained by a precessing jet
evolving in a dense circumstellar medium \citep{velazquez07}.

The OH emission at 1612, 1665, 1667, and 1720 MHz
\citep{aaquist93,miranda01,gomez09} is 
projected against the radio continuum structure, supporting the
association of these emissions.  Moreover, the OH maser emission at
1665 MHz arises 
within $\simeq 0\farcs 04$ from the peak of the radio continuum.
The spectrum of the OH maser emission at 1612 MHz shows at least four
features at $V_{\rm LSR}\simeq -$2.0, +8.8 (the brightest one), +18.4,
+21.2 km~s$^{-1}$.
The systemic velocity of this source is not well known:
while the velocity of the optical nebula is $\simeq 10$ km~s$^{-1}$ 
\citep{miranda00}, the velocity of the CO molecular gas is 
$\simeq 25$  km~s$^{-1}$ \citep{tafoya07}.
Interestingly, OH maser emission at 6035 MHz from this source has been recently
confirmed with interferometric observations \citep{desmurs10}. This
emission is very compact and located close in angular separation and
velocity to the 1720 MHz maser line \citep{gomez09}.  
This is the only young PN where the maser lines at 6035 and 1720 MHz have been confirmed until now.
 
In the VLA observations carried out by \citet{miranda01}, 
H$_2$O maser emission was also detected, within a region of $\simeq0\farcs 025$
from the peak of the radio continuum emission, as well as in two
regions 
at the tips of the bipolar radio jet $\simeq$1$''$ from the center. This
was the first PNe in which H$_2$O maser emission was confirmed. 
The kinematics of the water masers from
the central region suggests there is an
expanding and rotating disk, almost oriented perpendicular to the
innermost region of the observed jet, which may be related with the
collimation mechanism \citep{uscanga08}. 

\subsection{Possible OHPNe}
\label{possible}
In this category, on the one hand, we include sources in which 
the detection of OH maser emission has been carried out only with
single-dish observations, and therefore, interferometric observations are
needed to confirm the positional association with the radio continuum
emission of the source. On the other hand, we also consider those
sources here in which there is a confirmed association between maser and
continuum emission, but their nature as PNe has not yet been determined.
Possible OHPNe are listed in Table
\ref{possiblesources}.

\subsubsection{IRAS 17150$-$3754}
This source was listed by \citet{zijlstra89} as OHPN~2.
The radio continuum emission of this source was observed at 2 and 6~cm
with VLA by \citet{pottasch87}. The OH maser emission at 1612 MHz was
first detected in single-dish observations using the Parkes antenna
\citep{caswell81}, and later \citet{zijlstra89} confirmed that
detection using the VLA, only $\sim$1$''$ away from the continuum
emission. 
Our reanalysis of the VLA archive data
confirms the association of the OH and continuum positions (Table
\ref{possiblesources}). We have found that the OH spectrum at 1612 MHz 
shows a double peak, with features located at $V_{\rm LSR}\simeq -108.4$ km~s$^{-1}$ and $-103.4$ km~s$^{-1}$ and similar flux densities ($\simeq 280$ and 260 mJy, respectively).
It is listed as a possible PN in the Strasbourg-ESO catalog \citep{acker92}, but
its nature has not yet been  confirmed.

\subsubsection{IRAS 17168$-$3736}

This source is not listed in the OHPN catalog
by \citet{zijlstra89}, but it is considered as an OHPN  by
\citet{garciahernandez07}. 
The radio continuum emission of this source was observed at 20~cm with the VLA
by \citet{white05}, although it is not listed in their published catalog.
\citet{garciahernandez07} refer to a radio continuum position in 
the online version of the MAGPIS
catalog \citep{helfand06}. We also downloaded the
corresponding image from the MAGPIS catalog at 20~cm and confirm there is a
radio continuum source at the position given in Table
\ref{possiblesources}.

The OH maser emission was first detected in single-dish observations
using the Parkes antenna \citep{caswell81} and later,
\citet{sevenster97a} determined a more accurate interferometric 
position of the maser
emission at 1612 MHz, 
only $2\farcs 8$ away from the continuum emission,
both positions coinciding
within the errors (Table \ref{possiblesources}). 
The OH maser spectrum shows a double peak, with one feature 
located at $V_{\rm LSR}=-19.1$ km~s$^{-1}$ and the other at $+$7.2
km~s$^{-1}$. The intensity
ratio (blueshifted/redshifted) is about $\simeq 1.2$. 
This spectrum is similar to the one typically found in AGB
stars. 
The nature of this source is still uncertain. 
 
\subsubsection{IRAS 17221$-$3038}
This source was listed by \citet{zijlstra89} as OHPN~4.
The radio continuum emission of this source at 6~cm was detected by
\citet{pottasch88}.
We have confirmed this position in the review of the VLA archive
data
(Table  \ref{possiblesources}). This source was also detected at 20~cm in the NVSS \citep{condon98}. 
On the other hand, the reported OH maser emission at 1612 MHz has been detected
with the Nan\c{c}ay antenna \citep{zijlstra89}, and shows a double
peak with features separated $\simeq 33$ km~s$^{-1}$. 
Unfortunately, the
available VLA archive data 
in this source at 1612 MHz does not show any OH emission, since the
velocity range covered by these interferometric observations did not
include the OH features detected with the single-dish observations.

\subsubsection{IRAS 17375$-$2759}

This source  is listed by \citet{acker92} as a possible PN
and by \citet{zijlstra89} as OHPN~7.
The radio continuum emission from this source was detected at 6~cm with 
interferometric
observations by \citet{pottasch88}.
OH maser emission at 1612 MHz was detected in the ATCA/VLA survey
toward the Galactic bulge region \citep{sevenster97a}, only
$0\farcs 7$ away from this continuum emission (Table \ref{possiblesources}). Therefore both emissions are
associated considering the positional errors. The spectrum of the 1612
MHz OH line shows a double peak, with the spectral features 
at $V_{\rm
  LSR} =+23.2$ km~s$^{-1}$ and  +30.5 km~s$^{-1}$, and intensity ratio (blueshifted/redshifted) of $\simeq 3$. 
The nature of this
source is still uncertain.

\subsection{Sources previously misclassified as OHPNe}
\subsubsection{IRAS 17207$-$2856}
This source was listed by \citet{zijlstra89} as OHPN~3.
\citet{pottasch88} reported radio continuum emission with a flux
density of 8.2 mJy at 6~cm. However, in our analysis of archival data, we did not
detect any emission either at 3.6 or 6~cm, with upper limits of $\simeq 0.16$ mJy~beam$^{-1}$ (3$\sigma$) in
both cases. There is no radio continuum source at 20~cm in the NVSS catalog 
(3$\sigma$ upper limit $\simeq 1.4$ mJy~beam$^{-1}$).
Moreover, we could not confirm the presence of maser emission associated with this source in the VLA data archive, 
consistent with the VLA nondetection mentioned by 
\citet{zijlstra89}. 
This source had been
classified as an OHPN based on a single-dish detection of OH maser emission
with the Parkes antenna \citep{zijlstra89}, but it may arise from a
different source. In either case, the absence of radio continuum emission
allows us to reject it as an OHPN (see Table \ref{misclassified}).

\subsubsection{IRAS 17375$-$3000}
This source was listed by \citet{zijlstra89} as OHPN~8.
OH maser emission toward this source at 1612 MHz was detected with 
interferometric observations in the ACTA/VLA survey by
\citet{sevenster97a}. 
Its position is listed in Table \ref{misclassified}. 
The OH maser spectrum shows a double peak, with features located
at  $V_{\rm
  LSR} =-35.2$ km~s$^{-1}$ and $-$16.2 km~s$^{-1}$.
Radio continuum emission from this source could not be confirmed in
our reanalysis of two independent data sets from the VLA archive. 
Moreover, the radio continuum source
reported by \citet{ratag90} is $\simeq$20$''$ away
from the position of the OH maser. Considering that the beam of the
radio continuum observations of \citet{ratag90} ranged 
between $\simeq 5''$ and $10''$, and therefore, the
absolute positional uncertainty should be $\la 2''$,  we think it
is unlikely that the OH maser (with a positional error $\simeq$
2$\farcs$4) is associated with the reported continuum source, even if
the latter is real. 

\subsubsection{IRAS 17418$-$2713}
This source was considered as an OHPN by
\citet{garciahernandez07}, but it is not listed in the catalog of
\citet{zijlstra89}. However, the radio continuum emission at 6~cm 
toward this source reported by \citet{ratag91} is located 17$''$
away from the OH maser emission observed at 1612 MHz in the ACTA/VLA 
survey by \citet{sevenster97a}.
Its position is listed in Table \ref{misclassified}.
The OH maser spectrum shows a double peak, with features located at
 $V_{\rm
  LSR} = -27.9 $ km~s$^{-1}$
and  +2.8  km~s$^{-1}$.
The beam size of the radio continuum
observations was $\simeq$35$''$, and therefore the absolute positional
uncertainty should be $\la 7''$. Considering that the position
error of the OH maser is $\simeq$ 2$\farcs$4, we conclude that it is
not associated with the radio continuum emission reported by  \citet{ratag91}. 
In our reanalysis of VLA data we have found no radio continuum 
emission that may be associated with the OH maser.

Based on the infrared spectrum of this source that shows strong
amorphous silicate absorption features, together with high
variability, \citet{garciahernandez07} point out that this source
may still in the AGB phase, which also strongly suggests that 
this source is not an OHPN.
 
\subsubsection{IRAS 17443$-$2949}
This source was listed by \citet{zijlstra89} as OHPN~10.
Radio continuum emission toward this source was reported with the VLA
by \citet{ratag90}, while single-dish observations showed 
OH maser emission at 1612 MHz
\citep{zijlstra89}. 
However, \citet{gomez08} carried out simultaneous radio continuum and
OH maser observations toward this source. 
They detected both OH 1612 and 1665 MHz lines, but no 
radio continuum, either at 1.3 or 18~cm (Table \ref{misclassified}).  
No radio continuum emission was present in our reanalysis of the available
archival VLA data.
In any case, the radio continuum at 6~cm reported by \citet{ratag90}
is $\sim$10$''$ away from the position of the OH masers. Since this
separation is larger than the positional uncertainty ($\simeq 3''$),
we consider that the OH maser is not associated with the radio
continuum emission, even if the latter was real.

Moreover, \citet{gomez08} detected H$_2$O maser emission associated
with IRAS 17443$-$2949 located very close to the OH maser components
but also $\sim$9$''$ away from the radio continuum source reported by
\citet{ratag90}. 
On the other hand, 
\citet{garciahernandez07} suggest that this source is still in the
AGB phase. Based on its AGB nature, the detection of OH
and H$_2$O maser emission (the latter very rarely found in PNe), 
and its unlikely association with the radio
continuum emission,  \citet{gomez08} classified this source as an OH/IR
star. 
 
\subsubsection{IRAS 17580$-$3111}
This source was listed by \citet{zijlstra89} as OHPN~11.
\citet{ratag90} reported radio continuum emission at 6~cm toward this
source. OH maser emission at 1612 MHz was detected by
\citet{zijlstra89} in single-dish observations, showing at least four
spectral components.  
With interferometric VLA observations, \citet{gomez08}
detected OH maser emission at 1612 MHz with a typical double-peaked
spectrum, but radio continuum emission was not detected at either 1.3
or 18~cm (Table \ref{misclassified}). 
We could not detect any radio continuum emission in the
available VLA archival data. The position of the OH maser emission
(with an uncertainty 
of $\simeq 3''$) is 57$''$ away from the radio continuum source
reported by \citet{ratag90}, so is not associated with it,
even if the latter is real. 

In addition, \citet{gomez08} detected H$_2$O maser emission associated
with this source.  
\citet{garciahernandez07} classified IRAS 17580$-$3111 as a post-AGB
star, based on its infrared spectrum  and its low variability. 

\subsection{Related objects}
\label{related}
We have found other evolved 
objects that fulfill the observational
characteristics for being considered as OHPNe (showing both radio  continuum
and OH maser emission), but they are not PNe: e.g., 
symbiotic binary systems or post-AGB stars. Since our focus in this
paper is to search for true OHPNe, we have not carried out an
extensive search for these other objects. However, we mention a few
of them, to illustrate the importance of characterizing the
nature of evolved objects for a proper classification as OHPNe. We
also include them in the next sections, 
to compare their characteristics 
with those of bona fide OHPNe.

Two symbiotic systems are among the related objects: IRAS 17463$-$3700
\citep{aaquist90,ivison94} and IRAS 23412$-$1533 (R Aqr) \citep{kafatos89,ivison94},
as well as some post-AGB stars, such as IRAS 15367$-$5420, IRAS 15445$-$5449, and IRAS 16372$-$4808,
where the radio continuum and OH maser emissions are strongly associated, but in these sources the continuum emission may be arising from shocks since the central stars are still too cold to ionize their envelopes \citep{bains09}.
Particularly, in IRAS 15445$-$5449
water maser emission was detected 
with single-dish \citep{deacon07} and interferometric observations \citep{perez11}, with a wide velocity
spread. Therefore, this source is a ``water fountain'', a
type of object showing an early manifestation of collimated mass loss
in evolved stars \citep{imai07}.
We have included this interesting source in the following sections, for its relevance in the evolutionary scheme of intermediate-mass stars.

\section{General characteristics of OHPNe}
\label{characteristics}
Here we discuss some observational properties of the confirmed OHPNe
sources, 
comparing them with those of possible OHPNe and related
objects. 
We think that these observational
characteristics can be useful for future studies of these types of
objects, especially when more of them have been detected.

\subsection{Morphology}

\label{morphology}
All OHPNe for which the ionized emission has been resolved, present
a bipolar morphology at
optical, infrared and/or radio wavelengths. This is less clear in the case of
IRAS 19219$+$0947 (Vy 2$-$2) for which the HST images show an irregular
morphology \citep{sahai11}, although the internal kinematics do indicate
the presence of bipolar mass motions \citep{miranda91}. 
These bipolar morphologies
suggest that OH maser emission in PNe is related to nonspherical
mass-loss phenomena, rather than being the remnant of the masers
excited in the spherical AGB wind.
There is no resolved image of
the OHPN JaSt~23 and the four OHPN candidates (IRAS
17150$-$3702, IRAS 17168$-$3736, IRAS 17221$-$3038, and IRAS
17375$-$2759). 

Regarding the related objects, the symbiotic systems
IRAS 17463$-$3700 and IRAS 23412$-$1533 seem to show an
elongated structure at radio wavelengths \citep{kafatos89}. 
Moreover IRAS 23412$-$1533 (R Aqr) presents a clear bipolar 
morphology in optical images \citep{paresce94}.
The ``water fountain'' post-AGB star
IRAS 15445$-$5449 shows a bipolar structure with a dark lane 
in mid-infrared images \citep{lagadec11}. 

\subsection{Infrared colors}
\begin{figure}
\includegraphics[scale=0.43]{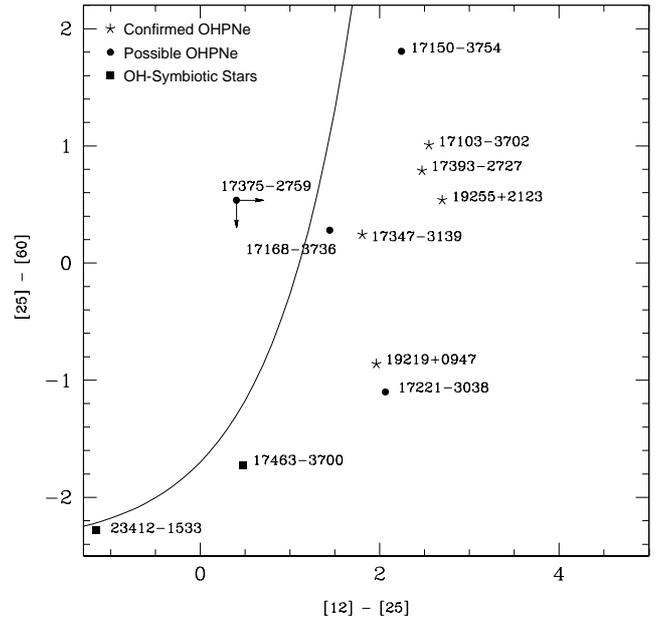}
\caption{IRAS two-color diagram with the position of the confirmed and possible OHPNe,
  as well as of related objects, with which IRAS data could be associated. Asterisks represent the confirmed OHPNe, circles are the candidate OHPNe, and squares are symbiotic stars. 
The solid line is the curve modeled by
  \citet{bedijn87} showing the location of AGB stars. 
 }
\label{iras2col}
\end{figure}

In Figure \ref{iras2col}, we present the IRAS two-color diagram of the confirmed OHPNe and related objects.
The IRAS colors are defined in the classical way as $[a]-[b]=-2.5$log$(S_a/S_b)$, where $S$ is the flux density at wavelengths $a$ and $b$ (in $\mu m$).
This diagram shows that all the confirmed OHPNe present a value of
$[12]-[25]> 1.0$, which is typical of PNe \citep{vanderveen88}, more
specifically between $\simeq 1.75$ and 2.75. The values of the
$[25]-[60]$ show a higher dispersion. 
The possible OHPNe present similar values for the  $[12]-[25]$ color. 
The far infrared colors of the symbiotic systems IRAS 17463$-$3700 and
IRAS 23412$-$1533 clearly differ from those of bona fide OHPN. They are
located in a region of the diagram commonly populated by variable
stars with young oxygen-rich circumstellar shells and AGB
stars \citep{vanderveen88}. Therefore, their infrared emission seem to be dominated by
the red giant companion. 

\begin{figure}[t!]
\includegraphics[scale=0.43]{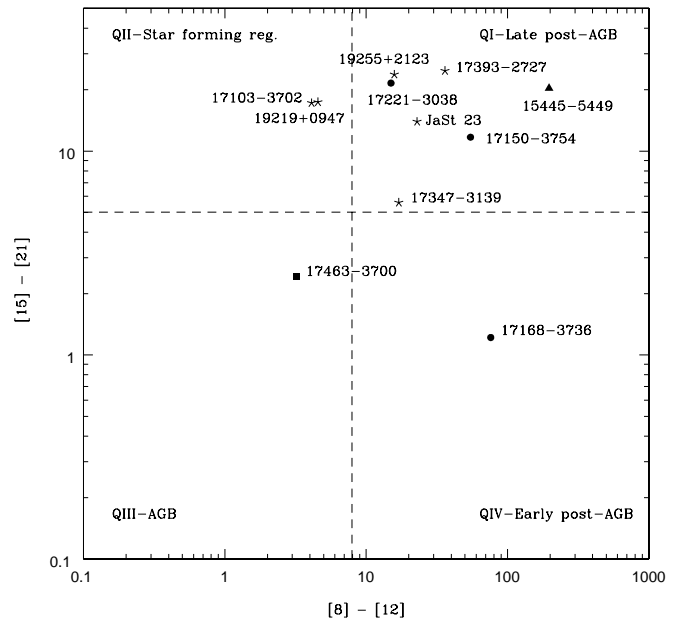}
\caption{MSX two-color diagram with the position of the confirmed and possible OHPNe and related objects for which MSX data are available. The vertical and horizontal lines divide the diagram into four quadrants that correspond to the areas of higher probability of containing the different types of objects labeled on the diagram \citep{sevenster02}. The symbols have the same meaning as in Fig. 1. The triangle represents the  ``water fountain'' post-AGB star IRAS 15445$-$5449.}
\label{msx2col}
\end{figure}

In Figure \ref{msx2col}, we show the position in the \mbox{[8]-[12]} $vs$
\mbox{[15]-[21]} two-color diagram of the confirmed OHPN sources and
related objects that have been observed with the Midcourse Space Experiment (MSX) satellite. 
\citet{sevenster02} defined four quadrants in this diagram, 
each of which tends to be populated by a different type of star.
Quadrants QI to
QIV tend to correspond with late
post-AGB stars, star-forming regions, AGB stars, and early post-AGB stars,
in that order. Although this diagram cannot be used to ascertain the
nature of a particular source \citep[e.g.,
  post-AGB stars and PNe can be found in QII and QIII,
  see][]{suarez04}, it is useful to see overall evolutionary trends.

Confirmed and possible OHPNe are mostly located in QI, consistent with
their proposed nature as young PNe. 
IRAS 17103$-$3702 and IRAS 19219$+$0947,
well-known PNe, 
are in QII,
although relatively close to the boundary. Their location may suggest
a differential characteristic with respect to the rest of sources, 
such as being in a more
evolved stage. On the other hand, the location of the candidate
IRAS 17168$-$3736 in QIV is well separated from the rest of sources,
which casts some doubt on its nature as an OHPN.

As a comparison, the symbiotic star 
IRAS 17463$-$3700 is located in QIII, also well-separated from the main
group of OHPNe.
Its location in QIII is consistent with the infrared emission being
dominated
by the red giant component of the system.  The post-AGB star IRAS
15445$-$5449 has colors similar to those of OHPNe, although it has
a higher value of [8]-[12].

\subsection{Spectral energy distributions}

We searched the Two Micron All Sky Survey (2MASS), Deep Near Infrared Survey of the Southern Sky (DENIS), MSX, IRAS, AKARI, Spitzer, Wide-field Infrared Survey Explorer (WISE), and Infrared Space Observatory (ISO) databases for infrared
photometric and spectroscopic data that could be used to construct the spectral energy
distributions (SEDs) of these objects. 
Combined data from these catalogs cover the spectral region from $\simeq$1 to
100 $\mu$m. 
The SEDs were constructed with these
infrared data, together with the available 
measurements of the radio continuum emission.

\begin{figure*}[!t]
\centering
\includegraphics[width=0.7\columnwidth]{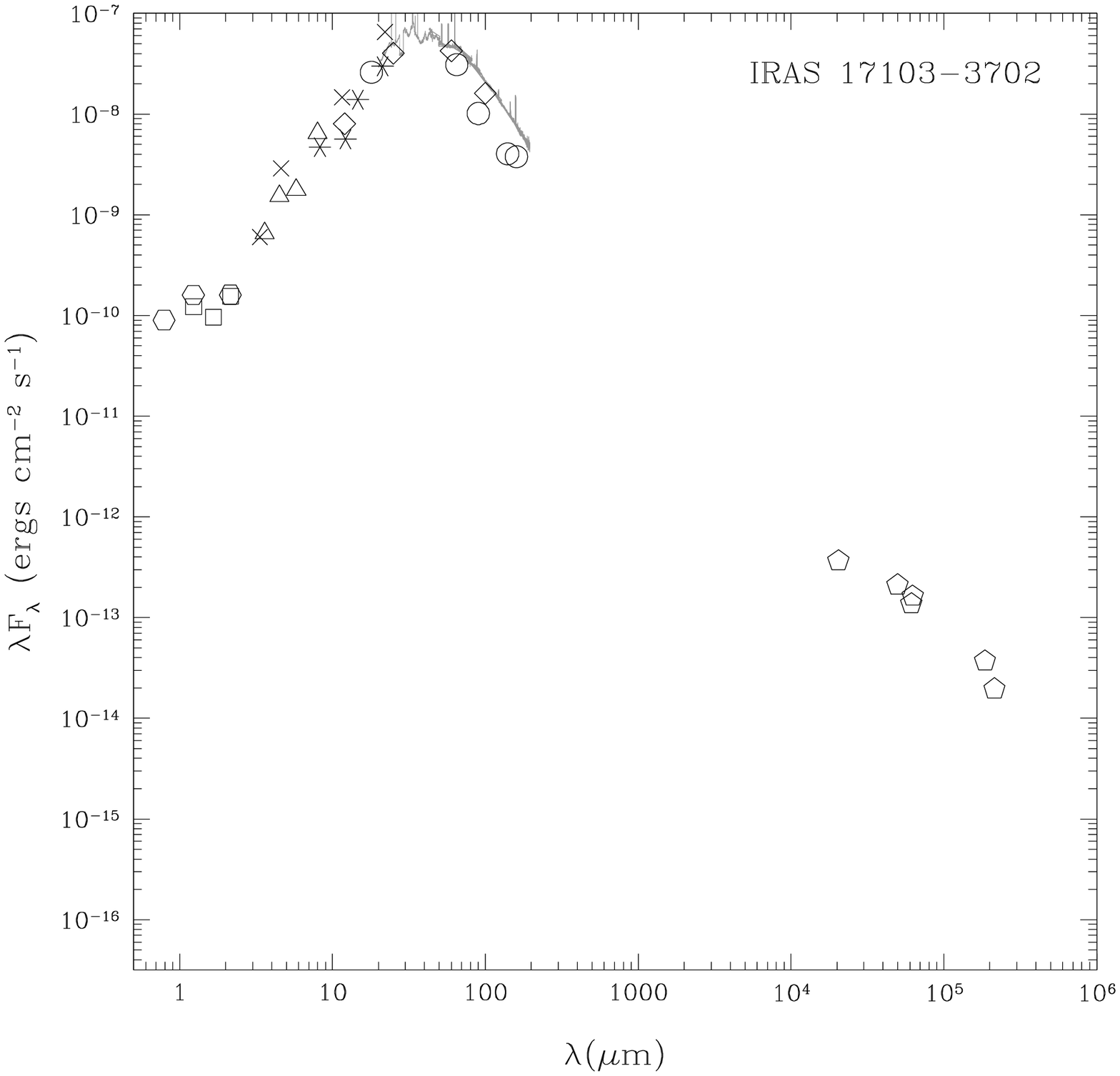}
\hspace*{\columnsep}%
\includegraphics[width=0.7\columnwidth]{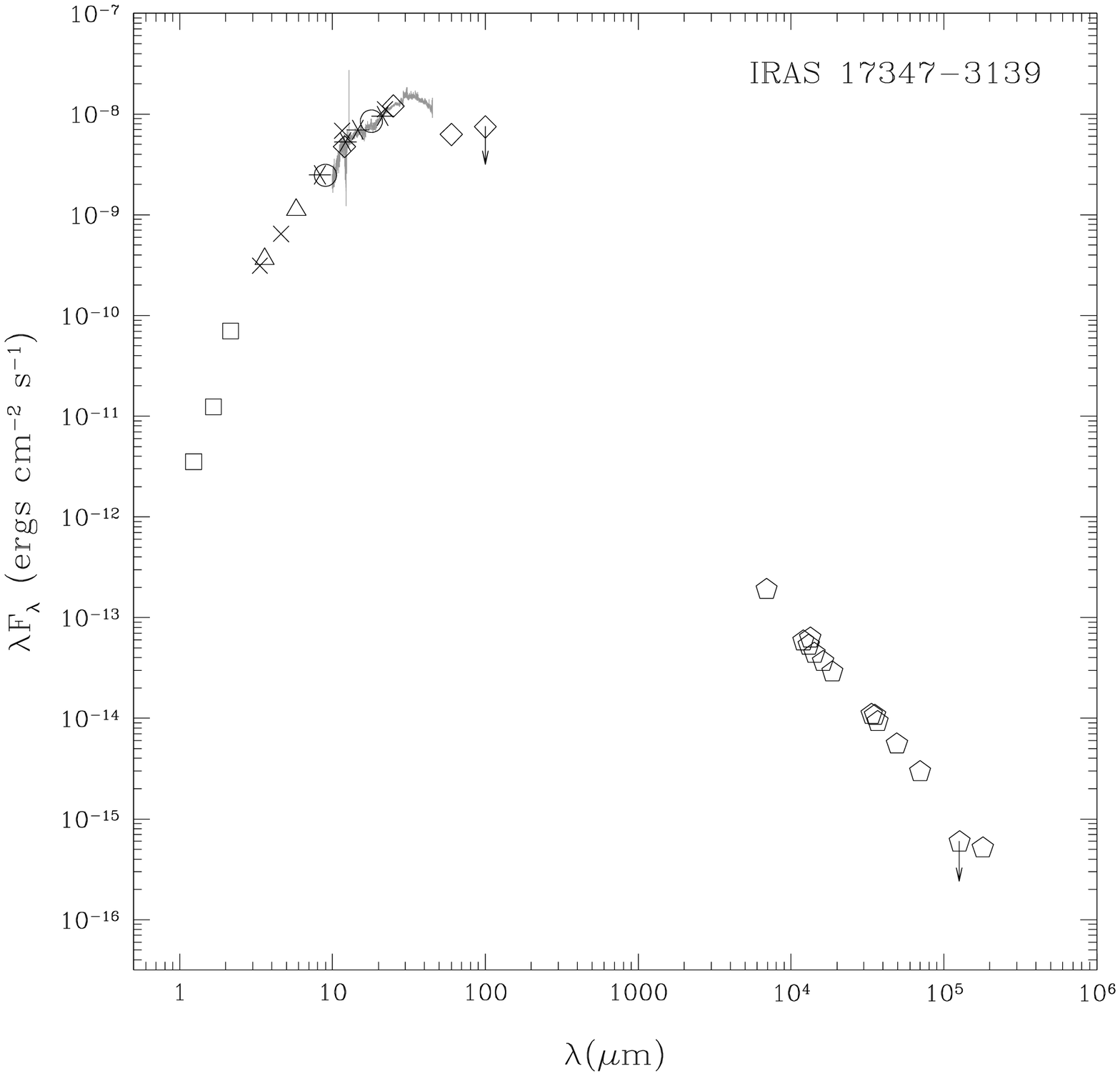}
\vskip .10in
\includegraphics[width=0.7\columnwidth]{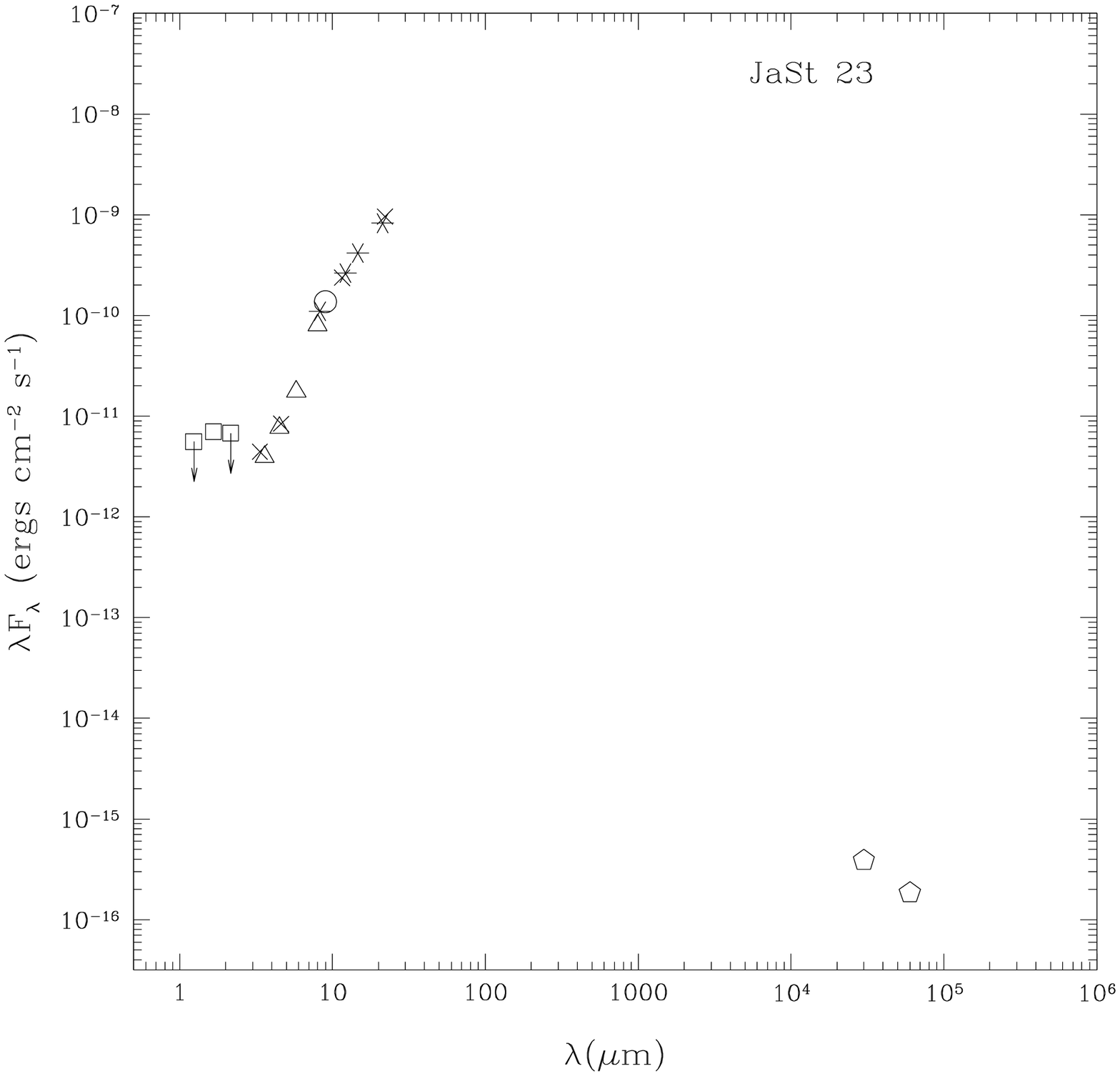}
\hspace*{\columnsep}%
\includegraphics[width=0.7\columnwidth]{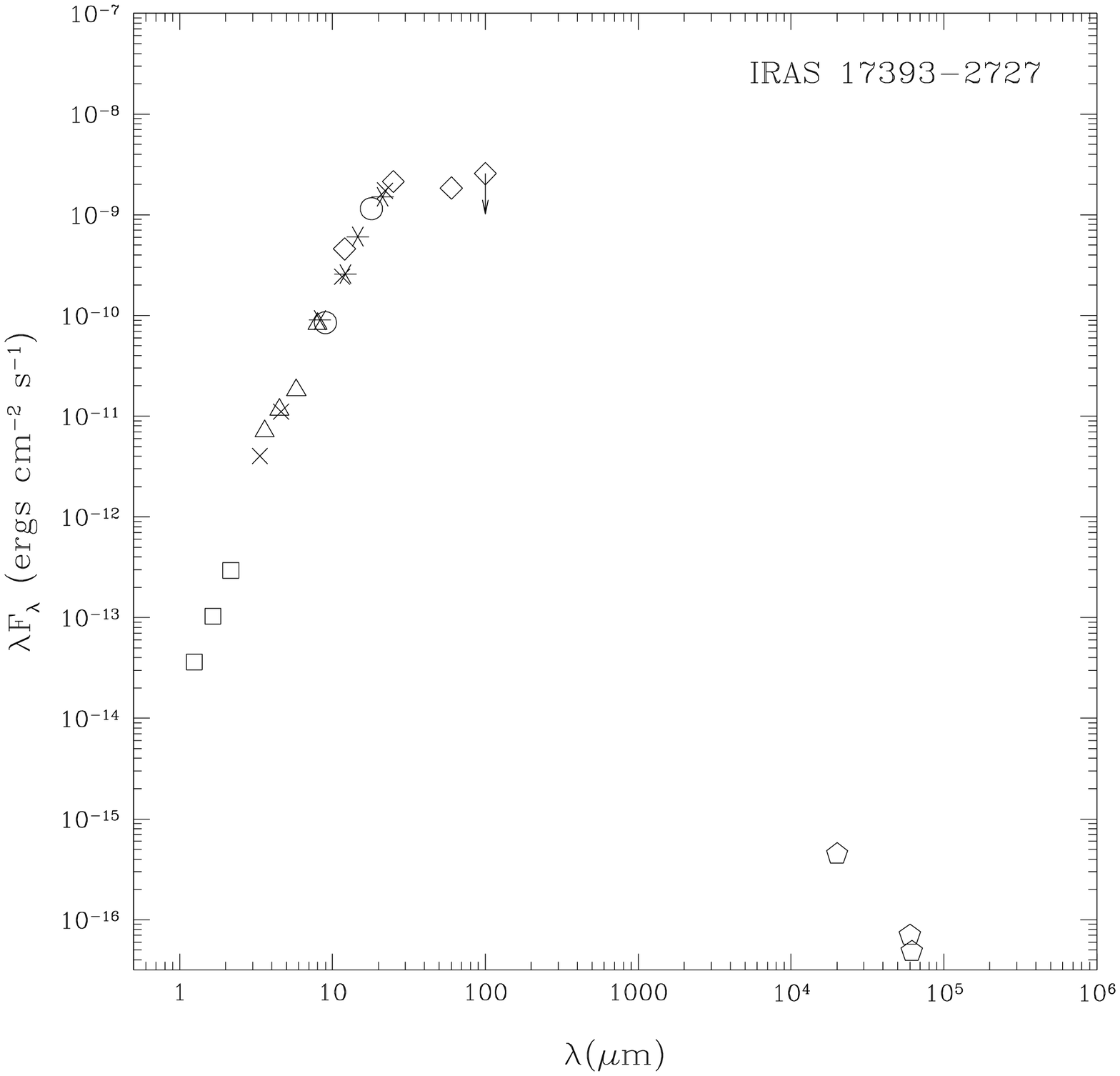}
\vskip .10in
\includegraphics[width=0.7\columnwidth]{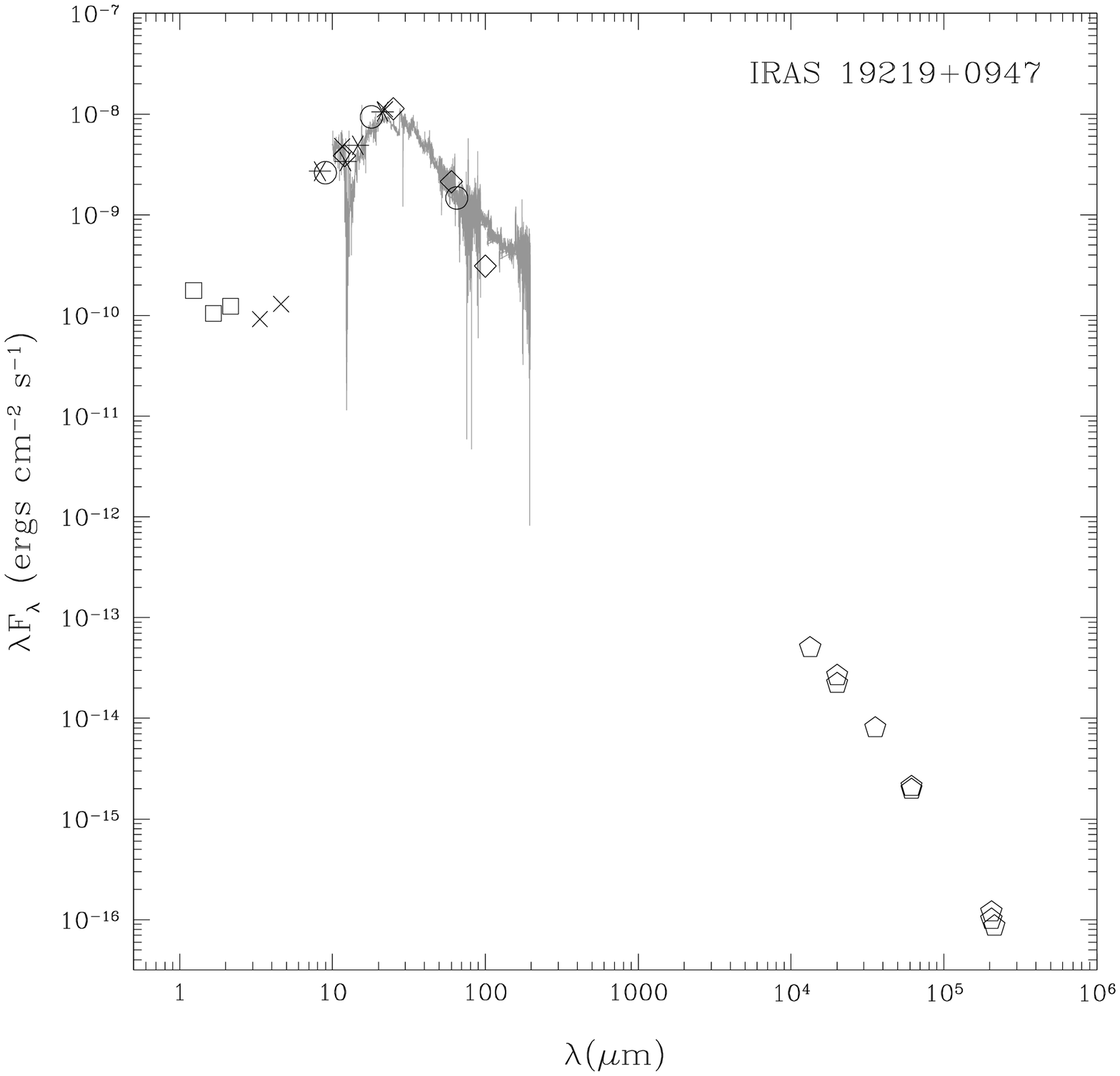}
\hspace*{\columnsep}%
\includegraphics[width=0.7\columnwidth]{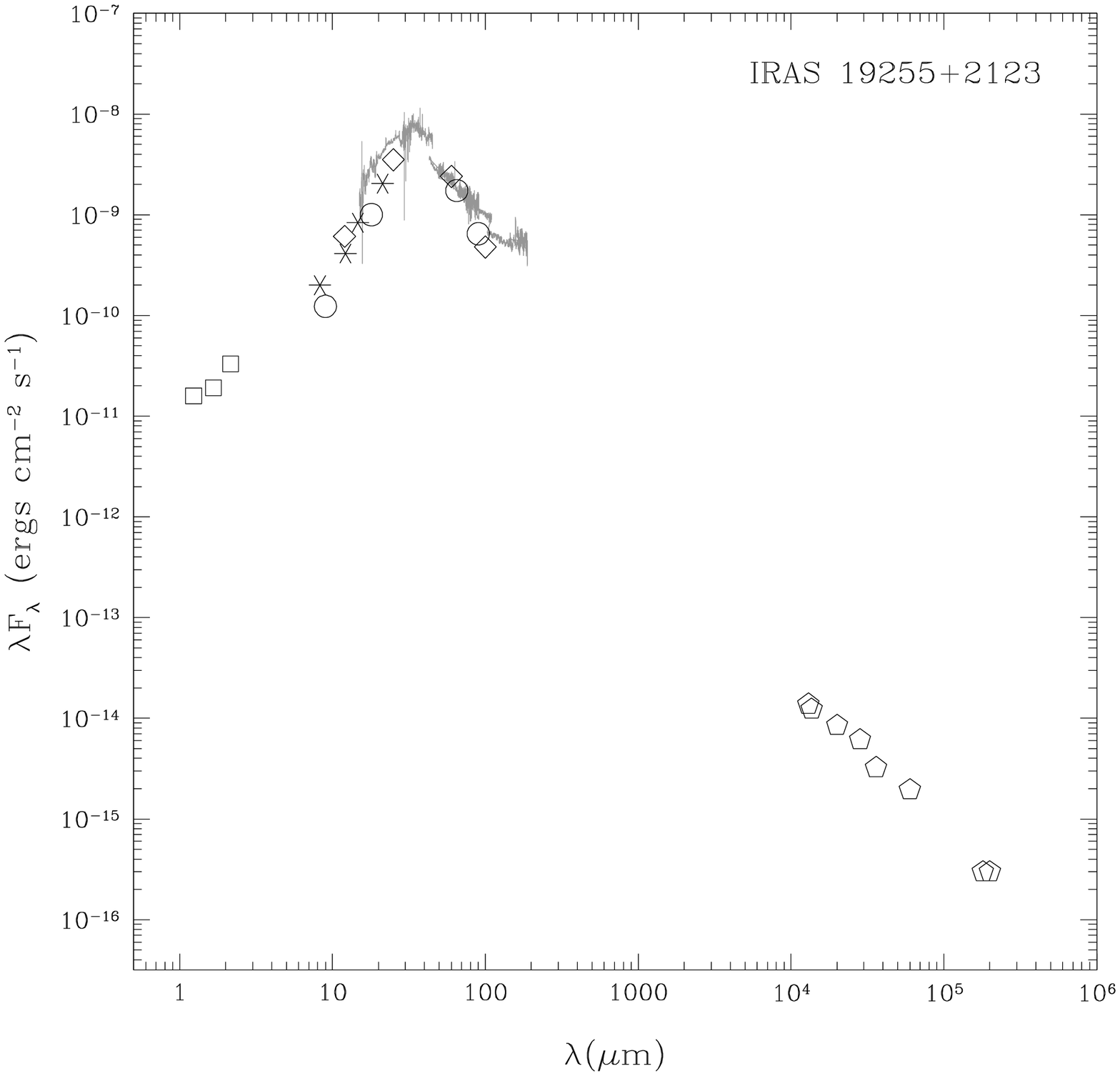}
\caption{SEDs of the confirmed OHPNe. The hexagons, squares, triangles, crosses, asterisks, circles, and diamonds represent the DENIS, 2MASS, Spitzer, WISE, MSX, Akari, and IRAS data, respectively. The pentagons represent measurements at different radio wavelengths. The ISO spectrum is shown in gray when available. The arrows indicate an upper limit value. 
The flux densities of radio continuum in the different sources were
measured by the following authors: IRAS 17103$-$3702
\citep{rodriguez85,payne88}, 17347$-$3139 \citep{gomez05,tafoya09},
JaSt23 \citep{vandsteene01}, IRAS 17393-2727 \citep[this
  paper]{pottasch87}, IRAS 19219$-$0947
\citep{seaquist83,christianto98,condon98}, and IRAS 19255+2123
\citep{aaquist91,miranda01,gomez09}. The Spitzer fluxes for IRAS 17103$-$3702 (triangles) and J, H, K fluxes (squares) 
for IRAS 17393-2727 were measured by \citet{phillips08} and \citet{ramoslarios12}, respectively. 
}
\label{sedconfirmed}
\end{figure*}

\begin{figure}[!t]
\centering
\includegraphics[width=0.7\columnwidth]{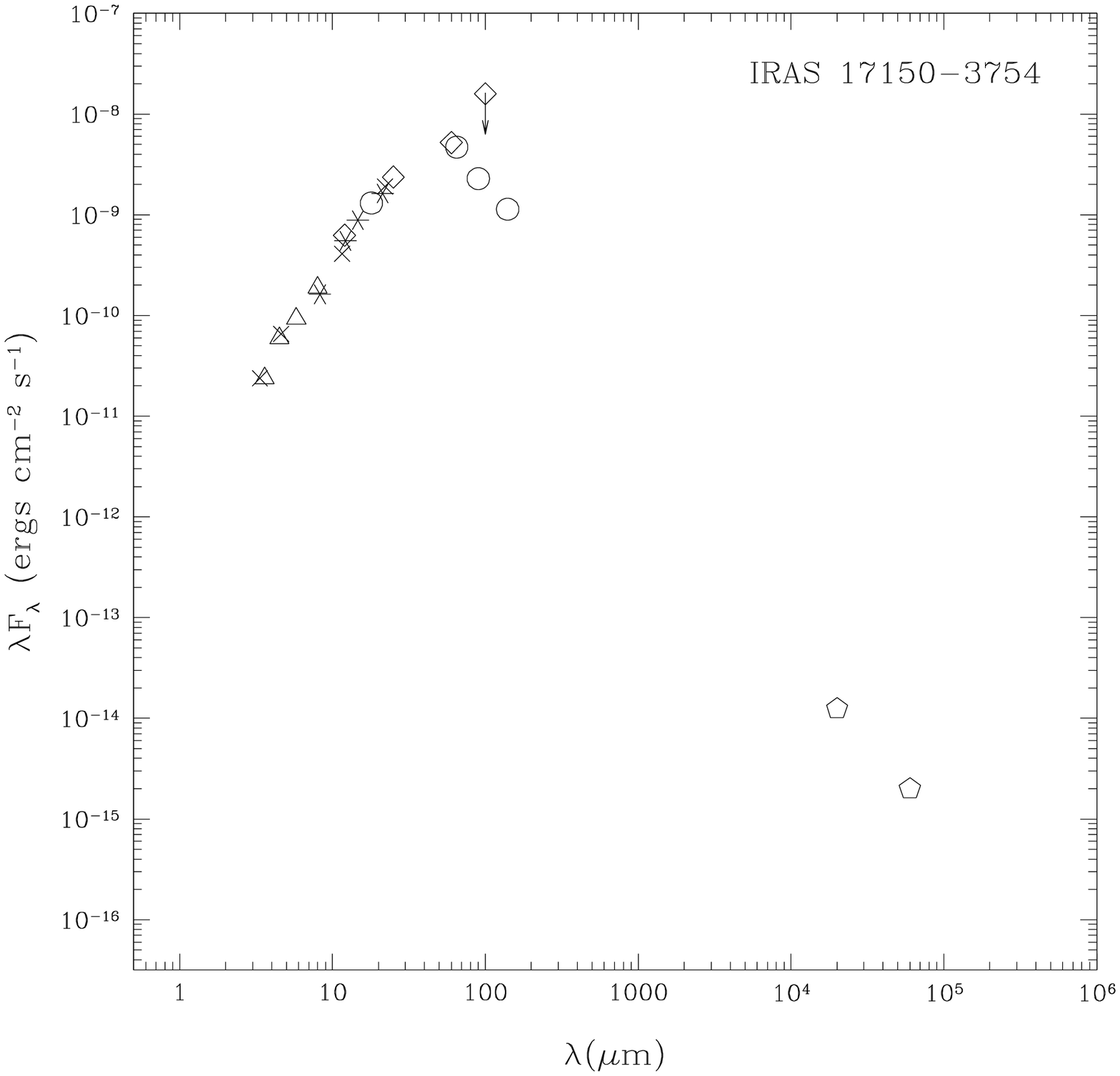}
\vskip .15in
\includegraphics[width=0.7\columnwidth]{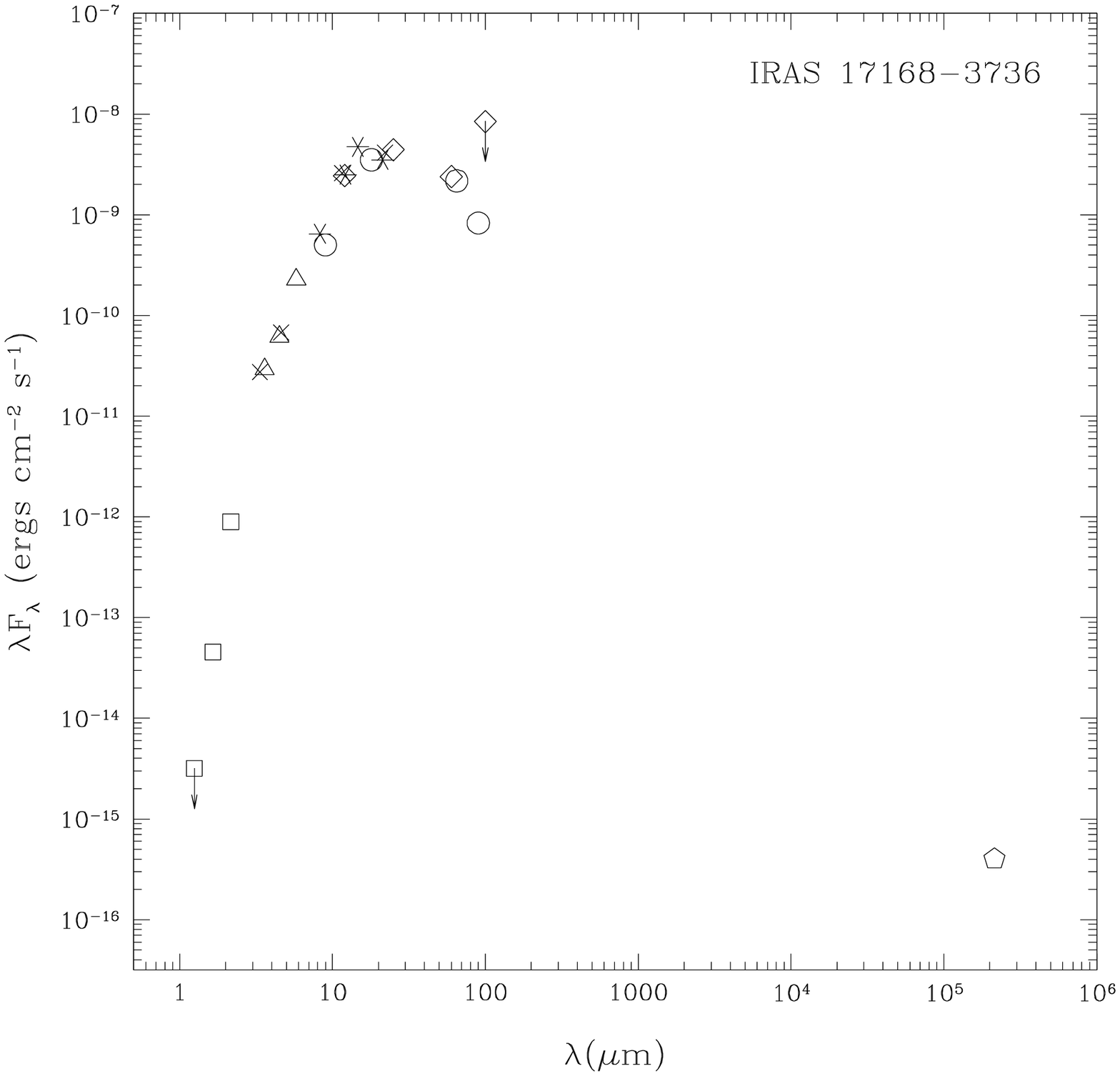}
\vskip .15in
\includegraphics[width=0.7\columnwidth]{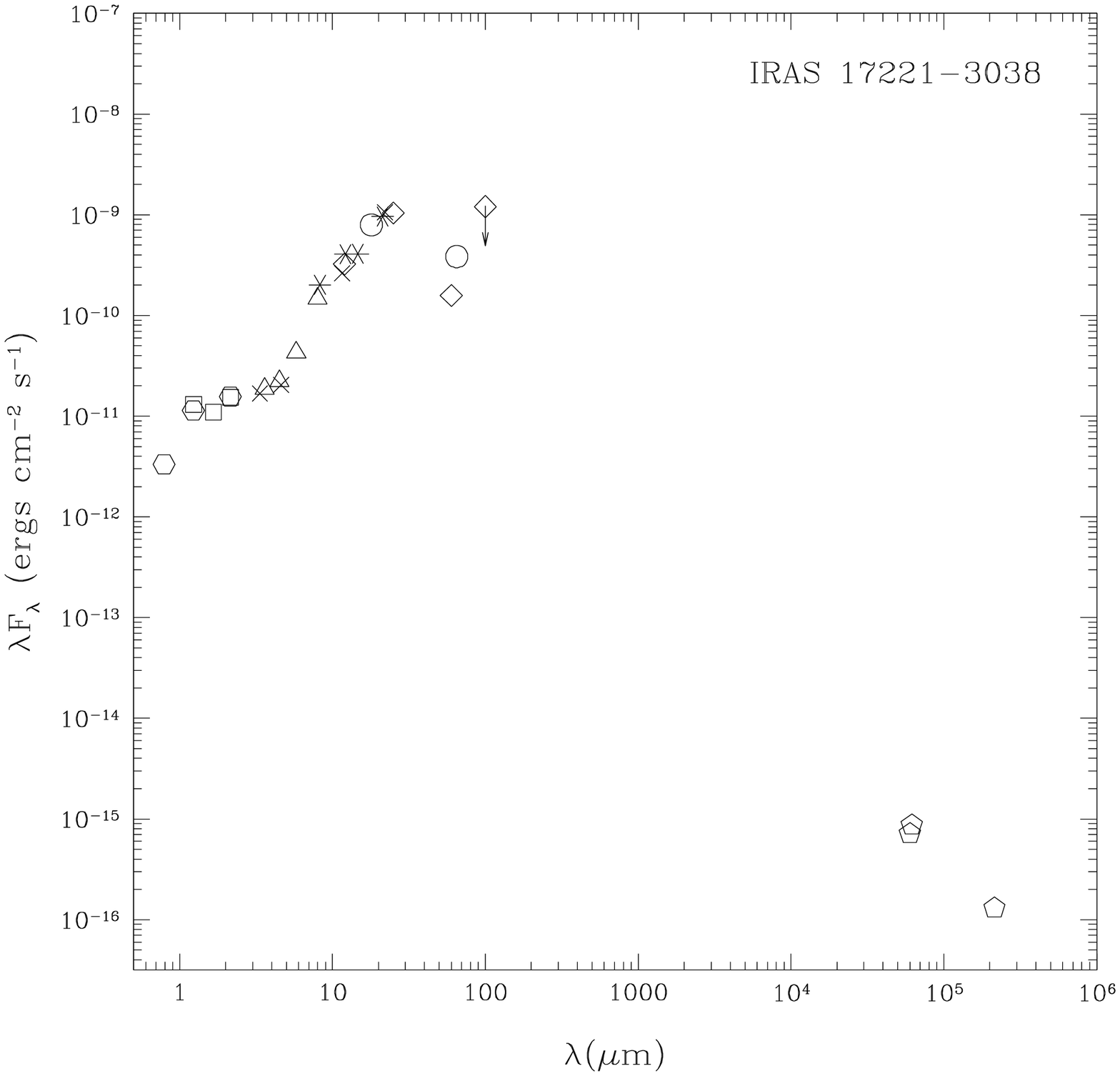}
\caption{SEDs of the possible OHPNe. Symbols mean the same as in Fig. 3. J, H, and K measurements taken by \citet{ramoslarios12}.
The flux densities of radio continuum in the different sources were
measured by the following authors: IRAS 17150$-$3754 \citep{pottasch87}, IRAS 17221$-$3038 \citep[this
  paper]{pottasch88,condon98}, and IRAS 17168$-$3736 (this paper).
}
\label{sedpossible}
\end{figure}

\begin{figure}[!t]
\centering
\includegraphics[width=0.7\columnwidth]{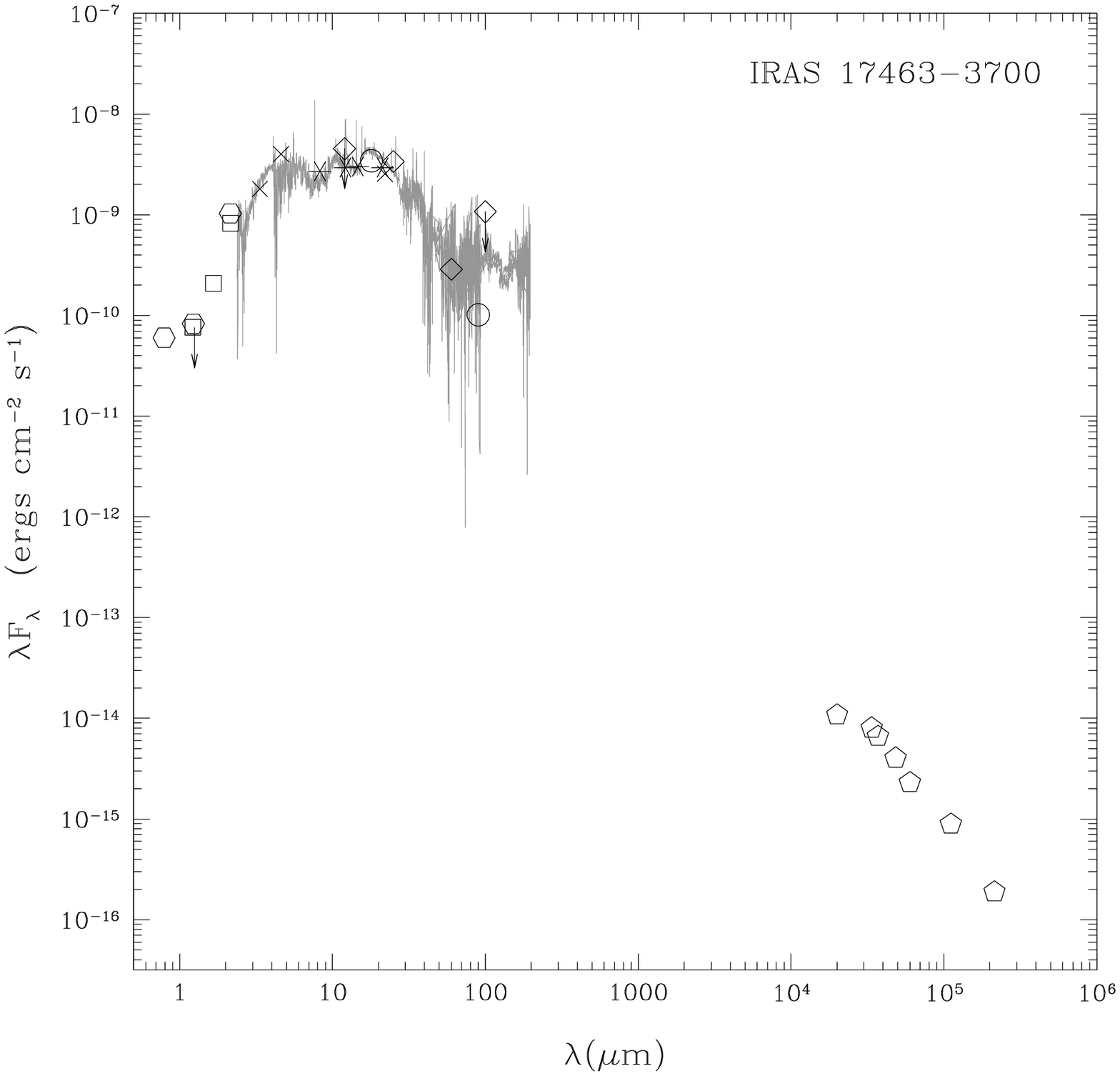}
\vskip .15in
\includegraphics[width=0.7\columnwidth]{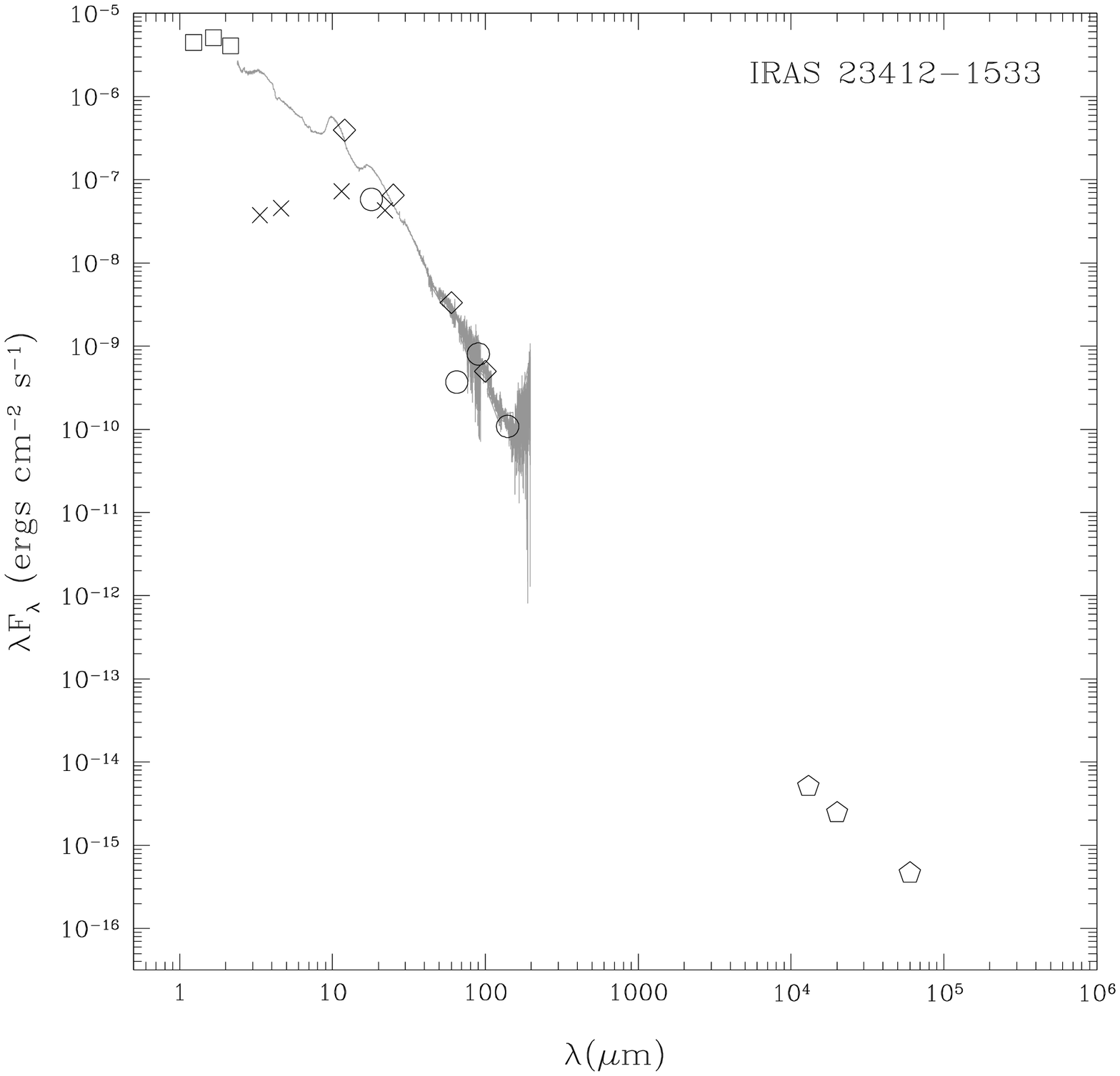}
\vskip .15in
\includegraphics[width=0.7\columnwidth]{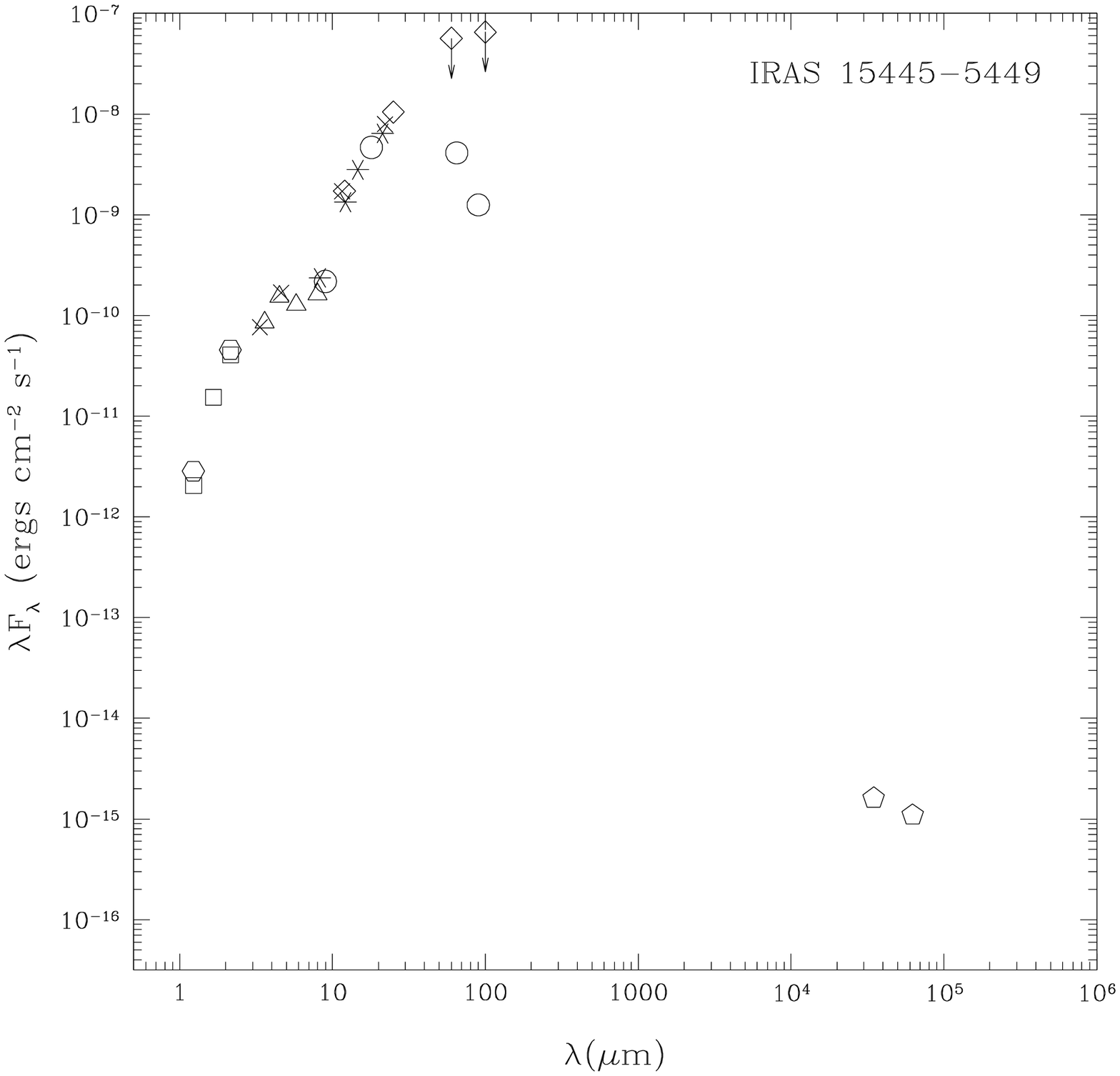}
\caption{SEDs of some of the related objects. Symbols mean the same as in Fig. 3. The ISO spectrum is shown in gray when available.  
The flux densities of radio continuum in the two sources were measured
by the following authors: IRAS 17463$-$3700
\citep{purton82,aaquist91,condon98}, IRAS 23412$-$1533
\citep{sopka82}, and IRAS 15445$-$5449 \citep{bains09}. 
}
\label{sedrelatedobj}
\end{figure}

Figure \ref{sedconfirmed} shows the SEDs of the confirmed OHPNe. All
show a clear peak at $\ga25$ $\mu$m (although the SED of  JaSt 23 is not well sampled at those
wavelengths), tracing the dust in the circumstellar envelope. 
Moreover, in the sources IRAS 17103$-$3702, IRAS
19219+0947, and JaSt 23, there is a hint that the emission flattens or rises again
at 
$\la 2$ $\mu$m, which could be the contribution from the central stellar
component or free-free emission from the ionized nebula 
\citep{zhang91}. 
In contrast, the SEDs of IRAS 17347$-$3139 and IRAS 17393$-$2727 present
a steep fall-off at $\lambda\la 25$ $\mu$m, and a more gradual one in 
IRAS 19255+2123, but no sign of the emission rising at short
wavelengths.
In these sources, the circumstellar envelope may still be
optically thick, obscuring the stellar and ionized component. They could be comparatively younger 
than IRAS 17103$-$3702, IRAS 19219+0947, and JaSt 23.

Figure \ref{sedpossible} shows the SEDs of the candidate OHPNe. We did not present the SED for 
IRAS 17375$-$2759, since the infrared counterpart of the radio emission is uncertain.
These candidate
sources also show a maximum at $\ga25$ $\mu$m. For the two sources that are well-sampled in the near infrared, we also see a dichotomy, with IRAS 17168$-$3736 showing a steep fall-off, while IRAS 17221$-$3038 seems to flatten at those wavelengths.

Finally, Figure \ref{sedrelatedobj} shows the SEDs of some related
objects (symbiotic and post-AGB stars showing both radio continuum and OH maser emission).  These SEDs are significantly
different from those of the OHPNe, in their shape and the location of the maximum. It seems that the SED
characteristics are useful for differentiating bona fide OHPNe from other OH-maser-emitting objects.

\subsection{Spectral properties of OH masers}
\label{spectraloh}

OH maser spectra in AGB stars typically present double-peaked
profiles, with narrow components separated by $\sim$30 km~s$^{-1}$, 
which have been interpreted as arising from the approaching
and receding sides of the circumstellar envelope \citep{sevenster97a}. The peaks are
sometimes asymmetrical, which could result in single-peaked
profiles if one of the peaks is below the sensitivity of the
instrument. However, the general trend is that the OH spectra are relatively symmetrical in those stars.
Departures from the double or single-peaked profile
are interpreted as tracing the departure from the nearly spherically
symmetric mass loss of the AGB \citep{deacon04}.

We note that spectra of all the confirmed OHPNe show strong asymmetry, preferentially with blueshifted emission with respect to the stellar velocity.
IRAS 17347$-$3139 and IRAS 17393$-$2727 show narrow blueshifted components, whose redshifted counterpart is absent or much weaker (almost by two orders of magnitude). IRAS 17103$-$3702, JaSt 23, and  IRAS 19219+0947 also show blueshifted emission with a relatively wide component ($\ga 10$ km~s$^{-1}$), while OH components in AGB stars are typically narrower. 
IRAS 19255+2123 shows four OH features, two redshifted and two blueshifted, but the brightest of them is blueshifted, assuming that the systemic velocity is given by the velocity of the optical nebula. We note that the four OH maser features are blueshifted with respect to the velocity of the molecular gas traced by CO and HCO$^+$, which is likely to arise from the circumstellar environment of the source \citep{tafoya07}.  
In all OHPNe, the OH spectra are different from the typical one in AGB stars. OHPN candidates also show strongly blueshifted and atypical spectra, with the exception of IRAS 17168$-$3736, which resembles an AGB spectrum. We note a possible observational bias, since one of the tools used by \citet{zijlstra89} to identify possible OHPNe was to search for radio continuum emission toward OH/IR stars whose redshifted OH emission was weak or absent. This criterion assumed that ionized gas could obscure OH emission in the background. However, among the sources in this paper, only the confirmed OHPN IRAS 17393$-$2727 and the candidate IRAS 17150$-$3754 were obtained with this observational bias. Therefore, the trend towards blueshifted spectra seems to be real and not an a priori bias.
In fact, blueshifted spectra can be due to two independent effects produced by the radio continuum emission: absorption of the redshifted OH emission and/or amplification of the radio continuum emission by the gas in the foreground (leading to increased blueshifted emission). 

The ionized gas in younger PNe tends to have larger emission measures \citep{kwok81b}, which translates into higher opacities of the free-free continuum emission, especially at low frequencies. For instance, in the case of IRAS 17347$-$3139, \citet{gomez05} determine an emission measure of $\simeq 1.5\times 10^{-9}$ cm$^{-6}$ pc. With this value, the optical depth of the radio continuum emission at 1612 MHz would be $\simeq 180$, and therefore the maser emission in the background would effectively be quenched. If the OH emission arises from an expanding envelope, this could explain a single blueshifted peak in this source. 

In addition to that, the presence of radio continuum emission could itself enhance the blueshifted OH emission. Maser emission occurs when the populations of molecules in the gas are inverted due to large energy input (e.g., via shocks or radiation). Background radiation going through this gas with inverted populations stimulates the emission, and it is exponentially amplified in the process. In the case of AGB stars, the background radiation that is amplified in OH masers can be the cosmic background or intrinsic spontaneous emission in the envelope. This background emission is weak, and a long optical path along gas with inverted population is needed. However, the free-free radiation provides a stronger background emission that can be amplified, leading to observable OH emission from a gas that would not be present without the radio continuum. This effect would produce maser emission from the foreground (approaching) gas.

\subsection{Relationship with H$_{2}$O maser-emitting PNe}

Water maser observations have been carried out toward the six confirmed OHPNe. Nondetections toward IRAS 17103$-$3702, IRAS 17393$-$2727, IRAS 19219+0947, and JaSt 23 have been reported by \citet{zuckerman87}, \citet{gomezy90}, \citet{degregorio04}, \citet{deacon07}, and \citet{suarez07}. On the other hand, detections of water masers have been confirmed through interferometric observations in IRAS 19255+2123 and IRAS 17347$-$3139 by \citet{miranda01} and \citet{degregorio04}, respectively. These are two of the three H$_2$O-maser-emitting PNe confirmed to date. Interestingly, the third H$_2$O-PNe already reported, IRAS 18061$-$2505 did not present any OH maser emission \citep{gomez08}, while more recent observations of IRAS 17347$-$3139 show that the H$_{2}$O masers have disappeared \citep{suarez07}.
In the case of the candidate OHPNe, IRAS 17150$-$3754 and IRAS 17168$-$3736 have also been observed, when searching for water maser emission, but nondetections were reported \citep{gomezy90,deacon07}. 

The presence of two objects in which masers from both species have been detected suggests a relationship 
between H$_{2}$O- and OH-emitting PNe. The relationship is also supported by the fact that PNe with either type of masers show bipolar morphologies \citep{miranda10}. It has been suggested that both H$_2$O-PNe and OHPNe originate in intermediate-mass stars, 5 to 8 M$_\odot$  \citep{suarez07, pottasch10}. It is possible that H$_2$O-PNe and OHPNe belong to the same underlying type of objects or that there is and evolutionary trend, with  H$_2$O-PNe being comparatively younger.
However, given the still small number of confirmed PNe with either H$_2$O or OH masers
and the variability of the maser emission, it is not possible to ascertain any such trend at this stage.

\section{Discussion and future prospects}
\label{discussion}

The confirmed OHPNe presented in this paper do not seem to constitute a homogeneous group, as they show significantly different characteristics. 
For example, IRAS 17347$-$3139 is optically obscured, and the PN JaSt 23 is very a compact source, suggesting that they are very young PNe. 
On the other hand, IRAS 17103$-$3702 and IRAS 19219+0947 show a complex structure in the optical, and they seem to be more evolved sources. In particular, IRAS 17103$-$3702 appears to be a full-blown PN.  
The different location in the MSX color-color diagram of these two sources (Figure \ref{msx2col}) also suggests they have distinctive characteristics. 

The presence or absence of different maser species has traditionally been proposed as a clock tracing different evolutionary stages \citep{lewis89,gomezy90}. In particular, masers were expected to sequentially disappear after the end of the AGB mass-loss, with timescales of $\simeq$10, 100, and 1000 yr for SiO, H$_2$O, and OH, respectively. This scheme implies that PNe showing OH maser emission must necessarily be very young. However, the existence of OHPNe in apparently different evolutionary stages suggests that this linear scheme of disappearance with time of the maser species may  not be applicable. Thus, the presence of OH may depend on other parameters than age, such as the mass-loss rate of the object or the particular orientation of the nebula with respect to the line of sight. 

The simple scheme of the sequential disappearance of masers is further challenged by the presence of water masers in ``water fountain" post-AGB stars \citep{imai07}, since these masers do not seem to be the remnants of the ones tracing the strong (and mostly spherical) mass loss in the AGB. In the case of water fountains, the masers trace highly collimated jets, which start in the late AGB or early post-AGB phase. In the case of PNe, the water masers pumped in the AGB are not expected to survive, given their short expected lifetimes (100 yr after the end of the AGB mass-loss), and yet water masers are present in PNe, which also indicate an excitation mechanism different from that in the AGB. Moreover, the existence of a PN with water masers \citep[IRAS 18061$-$2505, ][]{gomez08}, but without OH, does not easily fit a simple time-dependent scenario.

\citet{suarez09} propose an evolutionary scheme for water masers in evolved stars, in which this emission is first associated with the strong mass loss in the AGB, but when it ends, these masers can be excited by different processes, probably related to collimated mass loss in the post-AGB. These newly produced water masers would first trace jets (e.g. in water fountains) and disks in later stages (PNe). The case of OH masers could be different.
As discussed in Sect. \ref{spectraloh}, the presence of radio continuum emission could produce OH masers in physical conditions that were insufficient to produce maser emission before ionization started. It is then possible that OH masers pumped in the AGB phase disappear during the post-AGB phase, but reappear once the source becomes a PN, and its radio continuum emission is amplified by the OH molecules. Therefore, OH maser emission could last significantly longer than 1000 yr after the end of the AGB, even assuming that the maser-emitting material is the remnant of the one excited during that phase. The presence of OH masers in PN would be favored in sources with bright radio continuum emission, and therefore they could last longer in PNe with more massive central stars, which ionize a larger amount of gas in the envelope.

With the results of this paper, only six confirmed OHPNe and four candidate OHPNe have been identified. Finding a larger number of OHPNe will be an important step toward a  full understanding of the physical characteristics and evolution of these sources. This can be achieved with complete, sensitive, and unbiased surveys of radio continuum and OH masers. New instrumentation, such as ASKAP and APERTIF, are especially well-suited to large-scales surveys of OH maser and radio continuum emission, since their implementation of focal plane arrays (which increase the instantaneous field of view of radio interferometers) will provide a huge improvement in survey speed. In particular, the results from ASKAP surveys EMU \citep{norris11} in continuum and GASKAP \citep{dickey12} in OH, will help us compile a much more complete catalog of OHPNe in our galaxy.

\section{Summary}
\label{summary}

We have consulted the literature and public data archives, to obtain an updated catalog of 
OH-maser-emitting planetary nebulae (OHPNe), based on the spatial matching of sources found in
interferometric observations of radio continuum and OH maser emission. Our main conclusions are as follows.
\begin{itemize}
\item We have found six objects, previously identified as PNe, which show both radio continuum and OH maser emission. These can be considered as bona fide OHPNe. Four additional objects are candidate OHPNe, still to be confirmed.
\item All OHPNe for which the ionized emission has been resolved, present a bipolar morphology at optical, infrared, and/or radio wavelengths.
\item The infrared colors of OHPNe are consistent, with most of them being young PNe.
\item OH maser components in OHPNe tend to be blueshifted with respect to the systemic velocity. This could be due either to obscuration of the redshifted masers by optically thick radio continuum emission or to maser amplification of the radio continuum by the expanding gas closer to the observer.
\item OHPNe do not form an homogeneous group. Their characteristics indicate that they may span a wide range of relative ages. We suggest that OH masers pumped in the AGB phase may disappear during the post-AGB phase, but reappear once the source becomes a PN, and its radio continuum emission is amplified by the OH molecules. Therefore, OH maser emission could last significantly longer than 1000 yr after the end of the AGB, with longer lifetimes in PNe with more massive central stars.
\end{itemize}

\begin{acknowledgements}
LU and JFG wish to express their gratitude for the help and hospitality provided by 
the staff at CSIRO Science and Space Science (Australia) during part of the preparation of this paper.
This research has made use of the SIMBAD data base, operated at the CDS,
Strasbourg, France, and the NASA/IPAC Infrared Science Archive, which
is operated by the Jet Propulsion Laboratory, California Institute of
Technology, under contract with the National Aeronautics and Space
Administration. 
It makes use of data products from the Two Micron All Sky Survey (a joint project of the University of Massachusetts and the Infrared Processing and Analysis Center/California Institute of Technology, funded by NASA and NSF), DENIS (partly funded by the SCIENCE and the HCM plans of the European Commission under grants CT920791 and CT940627),
 Spitzer Space Telescope (operated by the Jet Propulsion Laboratory, California Institute of Technology under a contract with NASA),
WISE (a joint project of the University of California, Los Angeles, and the Jet Propulsion Laboratory/California Institute of Technology, funded by NASA), AKARI (a JAXA project with the participation of ESA), and MSX (funded by the Ballistic Missile Defense Organization with additional support from NASA Office of Space Science).
The authors acknowledge support from grants 
AYA2008-06189-C03-01 (JFG, OS, and LU) and
AYA 2011-30228-C03-01 (JFG, LFM, and OS) of the Spanish Ministerio de Ciencia e Innovaci\'on (MICINN), cofunded by FEDER funds. 
LFM is also partially supported by grants INCITE09E1R312096ES and
INCITE09312191PR of the Galician INCITE research program of the
Direcci\'on Xeral de Investigaci\'on, Desenvolvemento y Innovaci\'on of
the Spanish Xunta de Galicia. JFG acknowledges partial support from grant TIC-126 of Consejer\'{\i}a
de Econom\'{\i}a, Innovaci\'on y Ciencia of Junta de Andaluc\'{\i}a.

\end{acknowledgements}

\bibliographystyle{/home/lucero/OHPN/Articulo/bibtex/aa}

\begin{thebibliography}{33}

\expandafter\ifx\csname natexlab\endcsname\relax\def\natexlab#1{#1}\fi

\bibitem[{{Aaquist}(1993)}]{aaquist93} 
{Aaquist}, O.~B.\ 1993, \aap, 267, 260 

\bibitem[{{Aaquist} \& {Kwok}(1989)}]{aaquist89} 
{Aaquist}, O.~B., \& {Kwok}, S.\ 1989, \aap, 222, 227 

\bibitem[{{Aaquist} \& {Kwok}(1990)}]{aaquist90} 
{Aaquist}, O.~B., \& {Kwok}, S.\ 1990, \aaps, 84, 229 

\bibitem[{{Aaquist} \& {Kwok}(1991)}]{aaquist91} 
{Aaquist}, O.~B., \& {Kwok}, S.\ 1991, \apj, 378, 599 

\bibitem[Acker et al.(1992)]{acker92} Acker, A., Marcout, J., 
Ochsenbein, F., Stenholm, B., 
\& Tylenda, R.\ 1992, The
Strasbourg-ESO Catalogue of Galactic Planetary Nebulae (Garching:
European Southern Observatory) 

\bibitem[Bains et al.(2009)]{bains09} Bains, I., Cohen, M., 
Chapman, J.~M., Deacon, R.~M., \& Redman, M.~P.\ 2009, \mnras, 397,
1386 

\bibitem[Baudry \& Neri(2001)]{baudry01} Baudry, A., \& Neri,
  R.\ 2001, in Millimeter Interferometry, Proceedings from the IRAM
  Millimeter Interferometry Summer School 2, Ed. A. Dutrey,
  http://iram.fr/IRAMFR/IS/IS2002/html\_2/book.html, 20

\bibitem[{{Bedijn}(1987)}]{bedijn87} 
{Bedijn}, P.~J.\ 1987, \aap, 186, 136

\bibitem[{{Bl\"ocker}(1995)}]{blocker95}
{Bl\"ocker}, T. 1995, \aap, 299, 755

\bibitem[{{Boji{\v c}i{\'c}} {et al.}(2011){Boji{\v c}i{\'c}}, {Parker}, {Filipovi{\'c}}, \& {Frew}}]{bojicic11} 
{Boji{\v c}i{\'c}}, I.~S., {Parker}, Q.~A., {Filipovi{\'c}}, M.~D., 
\& {Frew}, D.~J.\ 2011, \mnras, 412, 223 

\bibitem[{{Bowers} {et~al.}(1989){Bowers}, {Johnston}, \& {de Vegt}}]{bowers89}
{Bowers}, P.~F., {Johnston}, K.~J., \& {de Vegt}, C. 1989, \apj, 340, 479

\bibitem[{{Briggs}(1995)}]{briggs95} 
{Briggs}, D.~S. 1995, PhD Thesis, New Mexico Institute of Mining and Technology

\bibitem[{{Caswell} {et al.}(1981){Caswell}, {Haynes}, {Goss}, \& {Mebold}}]{caswell81} 
{Caswell}, J.~L., {Haynes}, R.~F., {Goss}, W.~M., \& {Mebold}, U.\ 1981, Australian Journal of Physics, 34, 333 

\bibitem[{{Christianto} \& {Seaquist}(1998)}]{christianto98} 
{Christianto}, H., \& {Seaquist}, E.~R.\ 1998, \aj, 115, 2466 

\bibitem[Condon et al.(1998)]{condon98a} Condon, J.~J., Cotton, 
W.~D., Greisen, E.~W., et al.\ 1998, \aj, 115, 1693

\bibitem[{{Condon} \& {Kaplan}(1998)}]{condon98} 
{Condon}, J.~J., \& {Kaplan}, D.~L. 1998, \apjs, 117, 361 

\bibitem[{{Davis} {et al.}(1979){Davis}, {Seaquist}, \& {Purton}}]{davis79} 
{Davis}, L.~E., {Seaquist}, E.~R., \& {Purton}, C.~R.\ 1979, \apj, 230, 434 

\bibitem[Deacon et al.(2004)]{deacon04} 
Deacon, R.~M., Chapman, J.~M., \& Green, A.~J.\ 2004, \apjs, 155, 595 

\bibitem[Deacon et al.(2007)]{deacon07} 
Deacon, R.~M., Chapman, J.~M., Green, A.~J., \& Sevenster, M.~N.\ 2007, \apj, 658, 1096 

\bibitem[{de Gregorio-Monsalvo} {et al.}(2004){de Gregorio-Monsalvo}, {G{\'o}mez}, {Anglada}, {Cesaroni}, {Miranda}, {G{\'o}mez}, \& {Torrelles}]{degregorio04} 
{de Gregorio-Monsalvo}, I., {G{\'o}mez}, Y., {Anglada}, G., {Cesaroni}, R., {Miranda}, L.~F., {G{\'o}mez}, J.~F., \& {Torrelles}, J.~M.\ 2004, \apj, 601, 921 

\bibitem[Desmurs et al.(2002)]{desmurs02} 
Desmurs, J.-F., Baudry, A., Sivagnanam, P., \& Henkel, C.\ 2002, \aap, 394, 975 

\bibitem[Desmurs et al.(2010)]{desmurs10} 
Desmurs, J.-F., Baudry, A., Sivagnanam, P., et al.\ 2010, \aap, 520, A45 

\bibitem[Dickey et al.(2012)]{dickey12} 
Dickey, J.~M., McClure-Griffiths, N., Gibson, S.~J., et al.\ 2012,  PASA, in press

\bibitem[{{Engels} {et al.}(2010){Engels},{Bunzel}, \& {Heidmann}}]{engels10} 
{Engels}, D., {Bunzel}, F., \& {Heidmann}, B.\ 2010, Database of Circumstellar Masers v 2.0

\bibitem[{{Garc{\'{\i}}a-Hern{\'a}ndez}
  {et~al.}(2007){Garc{\'{\i}}a-Hern{\'a}ndez}, {Perea-Calder{\'o}n},
  {Bobrowsky}, \& {Garc{\'{\i}}a-Lario}}]{garciahernandez07}
{Garc{\'{\i}}a-Hern{\'a}ndez}, D.~A., {Perea-Calder{\'o}n}, J.~V., {Bobrowsky},
  M., \& {Garc{\'{\i}}a-Lario}, P. 2007, \apjl, 666, L33

\bibitem[{{G{\'o}mez} {et al.}(2005){G{\'o}mez}, {de 
Gregorio-Monsalvo}, {Lovell}, {Anglada}, {Miranda}, 
{Su{\'a}rez}, {Torrelles}, \& {G{\'o}mez, Y.}}]{gomez05} 
{G{\'o}mez}, J.~F., {de Gregorio-Monsalvo}, I., {Lovell}, J.~E.~J., {Anglada}, G., {Miranda}, L.~F., {Su{\'a}rez}, O., {Torrelles}, J.~M., \& {G{\'o}mez}, Y.\ 2005, \mnras, 364, 738 

\bibitem[{{G\'omez} {et~al.}(1990){G\'omez}, {Moran}, \&
  {Rodr\'{\i}guez}}]{gomezy90}
{G\'omez}, Y., {Moran}, J.~M., \& {Rodr\'{\i}guez}, L.~F. 1990, Revista
  Mexicana de Astronomia y Astrofisica, 20, 55

\bibitem[{{G\'omez} {et al.}(1989){G\'omez}, {Rodr\'{\i}guez}, {Moran}, \& {Garay}}]{gomez89} 
{G\'omez}, Y., {Rodr\'{\i}guez}, L.~F., {Moran}, J.~M., \& {Garay}, G.\ 1989, \apj, 345, 862 

\bibitem[{{G{\'o}mez} {et al.}(2008){G{\'o}mez}, {Su{\'a}rez}, {G{\'o}mez}, {Miranda}, {Torrelles}, {Anglada}, \& {Morata}}]{gomez08} 
G{\'o}mez, J.~F., 
Su{\'a}rez, O., G{\'o}mez, Y., Miranda, L.~F., Torrelles, J.~M., Anglada, 
G., \& Morata, {\'O}.\ 2008, \aj, 135, 2074 

\bibitem[{{G{\'o}mez} {et al.} (2009){G{\'o}mez}, {Tafoya}, {Anglada}, {Miranda}, {Torrelles}, {Patel}, 
\& {Hern{\'a}ndez}}]{gomez09} 
G{\'o}mez, Y., Tafoya, D., Anglada, G., Miranda, L.~F., Torrelles, J.~M., Patel, N.~A., 
\& Hern{\'a}ndez, R.~F.\ 2009, \apj, 695, 930 

\bibitem[Helfand et al.(2006)]{helfand06} Helfand, D.~J., Becker, 
R.~H., White, R.~L., Fallon, A., \& Tuttle, S.\ 2006, \aj, 131, 2525 

\bibitem[Imai(2007)]{imai07} 
Imai, H.\ 2007, in Astrophysical Masers and their Environments, IAU Symp. 242, 279 

\bibitem[{{Ivison} {et al.}(1994) {Ivison}, {Seaquist}, \& {Hall}}]{ivison94} 
{Ivison}, R.~J., {Seaquist}, E.~R., \& {Hall}, P.~J.\ 1994, \mnras, 269, 218 

\bibitem[{{Jacoby} \& {Van de Steene}(2004)}]{jacoby04} 
{Jacoby}, G.~H., \& {Van de Steene}, G.\ 2004, \aap, 419, 563 

\bibitem[Jewell et al.(1985)]{jewell85} 
Jewell, P.~R., Schenewerk, M.~S., \& Snyder, L.~E.\ 1985, \apj, 295, 183 

\bibitem[{{Kafatos} {et al.}(1989){Kafatos}, {Hollis}, {Yusef-Zadeh}, {Michalitsianos}, \& {Elitzur}}]{kafatos89}
{Kafatos}, M., {Hollis}, J.~M., {Yusef-Zadeh}, F., {Michalitsianos}, A.~G., 
\& {Elitzur}, M.\ 1989, \apj, 346, 991 

\bibitem[{{Kwok}(1993)}]{kwok93} 
{Kwok}, S. 1993, \araa, 31, 63 

\bibitem[{{Kwok} {et al.}(1981){Kwok}, {Purton}, \& {Keenan}}]{kwok81b} 
{Kwok}, S., {Purton}, C.~R., \& {Keenan}, D.~W.\ 1981, \apj, 250, 232 

\bibitem[Lagadec et al.(2011)]{lagadec11} Lagadec, E., Verhoelst, 
T., M{\'e}karnia, D., et al.\ 2011, \mnras, 1426 

\bibitem[{{Lewis}(1989)}]{lewis89}
{Lewis}, B.~M. 1989, \apj, 338, 234

\bibitem[{{Luo} {et al.}(2005){Luo}, {Condon}, \& {Yin}}]{luo05} 
{Luo}, S.~G., {Condon}, J.~J., \& {Yin}, Q.~F.\ 2005, \apjs, 159, 282 

\bibitem[{{Manteiga} {et al}.(2011){Manteiga}, {Garc{\'{\i}}a-Hern{\'a}ndez}, {Ulla}, {Manchado},\& {Garc{\'{\i}}a-Lario}}]{manteiga11} 
{Manteiga}, M., {Garc{\'{\i}}a-Hern{\'a}ndez}, D.~A., {Ulla}, A., {Manchado}, A., \& {Garc{\'{\i}}a-Lario}, P.\ 2011, \aj, 141, 80 

\bibitem[{{Meaburn} {et al.}(2005){Meaburn}, {L{\'o}pez}, {Steffen}, {Graham}, \& {Holloway}}]{meaburn05} 
Meaburn, J., L{\'o}pez, J.~A., Steffen, W., Graham, M.~F., \& Holloway, A.~J.\ 2005, \aj, 130, 2303 

\bibitem[{{Miranda} {et al.}(2000){Miranda}, {Fern{\'a}ndez}, {Alcal{\'a}}, {Guerrero}, {Anglada}, {G{\'o}mez}, {Torrelles}, \& {Aaquist}}]{miranda00} 
Miranda, L.~F., Fern{\'a}ndez, M., Alcal{\'a}, J.~M., Guerrero, M.~A., Anglada, G., G{\'o}mez, Y., Torrelles, J.~M., \& Aaquist, O.~B.\ 2000, \mnras, 311, 748 
\bibitem[{{Miranda} {et~al.}(2001){Miranda}, {G{\'o}mez}, {Anglada}, \&
  {Torrelles}}]{miranda01}
{Miranda}, L.~F., {G{\'o}mez}, Y., {Anglada}, G., \& {Torrelles}, J.~M. 2001,
  \nat, 414, 284

\bibitem[{{Miranda} \& {Solf}(1991)}]{miranda91} 
{Miranda}, L.~F., \& {Solf}, J.\ 1991, \aap, 252, 331 

\bibitem[{{Miranda} {et al.}(2010){Miranda},{ Su{\'a}rez}, \& {G{\'o}mez}}]{miranda10} 
{Miranda}, L.~F., {Su{\'a}rez}, O., \& {G{\'o}mez}, J.~F.\ 2010, Lecture Notes and Essays in Astrophysics, 4, 89 

\bibitem[Norris et al.(2011)]{norris11} 
Norris, R.~P., Hopkins, A.~M., Afonso, J., et al.\ 2011, \pasa, 28, 215 

\bibitem[Paresce \& Hack(1994)]{paresce94} Paresce, F., \& Hack, W.\ 1994, \aap, 287, 154 

\bibitem[{{Payne} {et al.}(1988){Payne}, {Phillips}, \& {Terzian}}]{payne88} 
{Payne}, H.~E., {Phillips}, J.~A., \& {Terzian}, Y.\ 1988, \apj, 326, 368

\bibitem[P{\'e}rez-S{\'a}nchez et al.(2011)]{perez11} 
P{\'e}rez-S{\'a}nchez, A.~F., Vlemmings, W.~H.~T., 
\& Chapman, J.~M.\ 2011, \mnras, 418, 1402 
 
\bibitem[Phillips \& Ramos-Larios (2008)]{phillips08} 
Phillips, J.~P., \& Ramos-Larios, G.\ 2008,  \mnras, 383, 1029 

\bibitem[Pottasch \& Bernard-Salas(2010)]{pottasch10} 
Pottasch, S.~R., \& Bernard-Salas, J.\ 2010, \aap, 517, A95 

\bibitem[{{Pottasch} {et~al.}(1987){Pottasch}, {Bignell}, \& {Zijlstra}}]{pottasch87} 
{Pottasch}, S.~R., {Bignell}, C., \& {Zijlstra}, A.\ 1987, \aap, 177, L49 

\bibitem[{{Pottasch} {et al.}(1988){Pottasch}, {Olling}, {Bignell}, \& {Zijlstra}}]{pottasch88} 
{Pottasch}, S.~R., {Olling}, R., {Bignell}, C., \& {Zijlstra}, A.~A.\ 1988, \aap, 205, 248 

\bibitem[Purton et al.(1982)]{purton82} 
Purton, C.~R., Feldman, P.~A., Marsh, K.~A., Allen, D.~A., \& Wright, A.~E.\ 1982, \mnras, 198, 321 

\bibitem[Ramos-Larios et al. (2012)]{ramoslarios12} 
Ramos-Larios et al.\ 2012, \aap, in press

\bibitem[{{Ratag} {et~al.}(1990){Ratag}, {Pottasch}, {Zijlstra}, \&
  {Menzies}}]{ratag90}
{Ratag}, M.~A., {Pottasch}, S.~R., {Zijlstra}, A.~A., \& {Menzies}, J. 1990,
  \aap, 233, 181

\bibitem[{{Ratag} \& {Pottasch}(1991)}]{ratag91} 
{Ratag}, M.~A., \& {Pottasch}, S.~R.\ 1991, \aaps, 91, 481 

\bibitem[{{Reid} \& {Moran}(1981)}]{reidmoran81}
{Reid}, M.~J. \& {Moran}, J.~M. 1981, \araa, 19, 231

\bibitem[{{Reid} {et~al.}(1977){Reid}, {Muhleman}, {Moran}, {Johnston}, \&
  {Schwartz}}]{reid77}
{Reid}, M.~J., {Muhleman}, D.~O., {Moran}, J.~M., {Johnston}, K.~J., \&
  {Schwartz}, P.~R. 1977, \apj, 214, 60

\bibitem[Reid et al.(1988)]{reid88} Reid, M.~J., Schneps, 
M.~H., Moran, J.~M., et al.\ 1988, \apj, 330, 809 

\bibitem[{{Rodr\'iguez} {et al.}(1985)}]{rodriguez85} 
{Rodr\'iguez}, L.~F., et al.\ 1985, \mnras, 215, 353 

\bibitem[{{Sahai} {et al}.(2007){Sahai}, {Morris}, {S{\'a}nchez-Contreras}, \& {Claussen}}]{sahai07} 
{Sahai}, R., {Morris}, M., {S{\'a}nchez-Contreras}, C., \& {Claussen}, M.\ 2007, \aj, 134, 2200 

\bibitem[{{Sahai} {et al.}(2011){Sahai}, {Morris}, \& {Villar}}]{sahai11} 
{Sahai}, R., {Morris}, M.~R., \& {Villar}, G.~G.\ 2011, \aj, 141, 134 

\bibitem[Schneider et al.(1983)]{schneider83} Schneider, S.~E., 
Terzian, Y., Purgathofer, A., \& Perinotto, M.\ 1983, \apjs, 52, 399 

\bibitem[{{Seaquist} \& {Davis}(1983)}]{seaquist83} 
{Seaquist}, E.~R., \& {Davis}, L.~E.\ 1983, \apj, 274, 659 

\bibitem[Seaquist(1991)]{seaquist91} Seaquist, E.~R.\ 1991, \aj, 
101, 2141 

\bibitem[{{Sevenster}(2002)}]{sevenster02} 
{Sevenster}, M.~N.\ 2002, \aj, 123, 2772 

\bibitem[{{Sevenster} {et~al.}(1997a){Sevenster}, {Chapman}, {Habing}, {Killeen}, \& {Lindqvist}}]{sevenster97a} 
{Sevenster}, M.~N., {Chapman}, J.~M., {Habing}, H.~J., {Killeen}, N.~E.~B., \& {Lindqvist}, M.\ 1997a, \aaps, 122, 79

\bibitem[{{Sevenster} {et~al.}(1997b){Sevenster}, {Chapman}, {Habing}, {Killeen}, \& {Lindqvist}}]{sevenster97b} 
{Sevenster}, M.~N., {Chapman}, J.~M., {Habing}, H.~J., {Killeen}, N.~E.~B., \& {Lindqvist}, M.\ 1997b, \aaps, 124, 509 

\bibitem[{{Sevenster} {et~al.}(2001){Sevenster},{van Langevelde}, {Moody}, {Chapman}, {Habing}, \& {Killeen}}]{sevenster01} 
{Sevenster}, M.~N., {van Langevelde}, H.~J., {Moody}, R.~A., {Chapman}, J.~M., {Habing}, H.~J., \& {Killeen}, N.~E.~B.\ 2001, \aap, 366, 481 

\bibitem[Shepherd et al.(1990)]{shepherd90} Shepherd, M.~C., 
Cohen, R.~J., Gaylard, M.~J., \& West, M.~E.\ 1990, \nat, 344, 522 

\bibitem[Sopka et al.(1982)]{sopka82} Sopka, R.~J., Herbig, G., 
Kafatos, M., \& Michalitsianos, A.~G.\ 1982, \apjl, 258, L35 

\bibitem[{{Su\'arez}(2004)}]{suarez04} 
{Su\'arez}, O.\ 2004, Ph.D. Thesis, Universidad de Vigo, Spain 

\bibitem[Su{\'a}rez et al.(2009)]{suarez09} 
Su{\'a}rez, O., G{\'o}mez, J.~F., Miranda, L.~F., et al.\ 2009, \aap, 505, 217 

\bibitem[Su{\'a}rez et al.(2007)]{suarez07} 
Su{\'a}rez, O., G{\'o}mez, J.~F., \& Morata, O.\ 2007, \aap, 467, 1085 

\bibitem[Tafoya et al.(2007)]{tafoya07} Tafoya, D., G{\'o}mez, 
Y., Anglada, G., et al.\ 2007, \aj, 133, 364 

\bibitem[{{Tafoya} {et~al.}(2009){Tafoya}, {G{\'o}mez}, {Patel},{Torrelles}, {G{\'o}mez}, {Anglada}, {Miranda}, \& {de Gregorio-Monsalvo}}]{tafoya09} 
{Tafoya}, D., {G{\'o}mez}, Y., {Patel}, N.~A., {et~al.} 2009, \apj, 691, 611 

\bibitem[Uscanga et al.(2008)]{uscanga08} 
Uscanga, L., G{\'o}mez, Y., Raga, A.~C., et al.\ 2008, \mnras, 390, 1127 

\bibitem[{{Van de Steene} \& {Jacoby}(2001)}]{vandsteene01} 
{Van de Steene}, G.~C., \& {Jacoby}, G.~H.\ 2001, \aap, 373, 536 

\bibitem[{{Van de Steene} \& {Pottasch}(1993)}]{vandsteene93}
{Van de Steene}, G.~C.~M. \& {Pottasch}, S.~R. 1993, \aap, 274, 895

\bibitem[{{Van de Steene} \& {Pottasch}(1995)}]{vandsteene95} 
{Van de Steene}, G.~C., \& {Pottasch}, S.~R.\ 1995, \aap, 299, 238 

\bibitem[{{van der Veen} \& {Habing}(1988)}]{vanderveen88} 
{van der Veen}, W.~E.~C.~J., \& {Habing}, H.~J.\ 1988, \aap, 194, 125 

\bibitem[Vel{\'a}zquez et al.(2007)]{velazquez07} Vel{\'a}zquez, 
P.~F., G{\'o}mez, Y., Esquivel, A., \& Raga, A.~C.\ 2007, \mnras, 382, 1965 

\bibitem[{{White, Becker,} \& {Helfand}(2005)}]{white05} 
{White}, R.~L., {Becker}, R.~H., \& {Helfand}, D.~J.\ 2005, \aj, 130, 586 

\bibitem[{{Zhang} \& {Kwok}(1991)}]{zhang91} 
{Zhang}, C.~Y., \& {Kwok}, S.\ 1991, \aap, 250, 179 

\bibitem[Zijlstra et al.(1989)]{zijlstra89} 
Zijlstra, A.~A., te Lintel Hekkert, P., Pottasch, S.~R., et al.\ 1989, \aap, 217, 157 

\bibitem[Zuckerman \& Lo(1987)]{zuckerman87} 
Zuckerman, B., \& Lo, K.~Y.\ 1987, \aap, 173, 263 


\end{thebibliography}

\newpage
\begin{table}
\begin{center}
\caption{VLA archival data used \label{archivaldata}}
\begin{tabular}{lllll}
\hline \hline 
Project ID & Config\tablefootmark{a} & Frequency & Type\tablefootmark{b} & Observing Date \\
           &                         & MHz       \\
\hline
SEAQ       & A                       & 1465      & cont                  & 1982-FEB-82 \\
           & A                       & 4885      & cont                  & 1982-FEB-82 \\  
AP076      & C                       & 4860      & cont                  & 1984-APR-23 \\
           & C                       & 4860      & cont                  & 1984-APR-24 \\
AP116      & B                       & 4860      & cont                  & 1986-JUL-26 \\
AP121      & BC                      & 14940     & cont                  & 1986-SEP-17 \\
AP125      & C                       & 4860      & cont                  & 1986-NOV-14 \\
           & C                       & 4860      & cont                  & 1986-DEC-11 \\
AP128      & C                       & 4860      & cont                  & 1987-JAN-17 \\
AS280      & A                       & 22460     & cont                  & 1987-JUL-21 \\
AP166      & A                       & 8440      & cont                  & 1988-NOV-06 \\
VFB01      & BC                      & 1615      & cont                  & 1990-OCT-18 \\
SEAQ       & A                       & 1612      & OH                    & 1982-FEB-82 \\
AH100      & BC                      & 1612      & OH                    & 1985-JUN-23 \\
AP163      & D                       & 1612      & OH                    & 1988-AUG-08 \\
AH185      & A                       & 1612      & OH                    & 1988-OCT-30 \\
AZ030      & D                       & 1612      & OH                    & 1987-MAR-25 \\
AP128      & C                       & 1665      & OH                    & 1987-JAN-17 \\
\hline
\end{tabular}
\end{center}
\tablefoottext{a}{Configuration of the VLA}
\tablefoottext{b}{Type of observation: cont (continuum), OH (OH maser)}
\end{table}

\begin{landscape}
\begin{table}
\begin{center}
\caption{Confirmed OHPNe \label{confirmations}}
\begin{tabular}{clllllllllccc}
\hline \hline 
Object ID & IRAS & PN G \tablefootmark{a} & Common name  &
\multicolumn{2}{c}{Continuum position} & Error \tablefootmark{b} &
\multicolumn{2}{c}{OH position}  & Error\tablefootmark{b} & Position
difference & References\tablefootmark{c} & Notes \\
  & & & & RA (J2000) & Dec (J2000)  & ($''$) & RA (J2000) & Dec (J2000)  &   ($''$) &($''$) &        \\
\hline
1 & 17103$-$3702  & 349.5+01.0 & NGC 6302 & 17 13 44.5 & $-$37 06 11 & 3 & 17 13 44.40 & $-$37 06 09.6 & 3 & 2 & 6 & \tablefootmark{d} \\
2 & 17347$-$3139 &&& 17 38 00.61 & $-$31 40 55.0 & 0.8 & 17 38 00.57 & $-$31 40 54.9 & 0.8 & 0.5 & 4  & \tablefootmark{d}  \\
3 & && JaSt 23& 17 40 23.08 & $-$27 49 12.3 & 0.4 & 17 40 23.07 & $-$27 49 11.4 & 2.4 & 0.9 & 2,3,6 &  \tablefootmark{e} \\  
4 & 17393$-$2727 &&& 17 42 33.14 & $-$27 28 24.7 & 0.8 & 17 42 33.16 & $-$27 28 24.6 & 2.4  & 0.3 & 1, 2, 6 \\ 
5 & 19219$+$0947 & 045.4$-$02.7 & Vy 2$-$2 & 19 24 22.218 & $+$09 53 56.33 & 0.09 & 19 24 22.207 & $+$09 53 56.14 & 0.22 & 0.25 & 6 &   \\
6 & 19255$+$2123 & 056.0+02.0 & K 3$-$35 & 19 27 44.026 & $+$21 30 03.57 & 0.20 & 19 27 44.022 &    $+$21 30 03.31 & 0.20 & 0.2 & 5 & \tablefootmark{d} \\      
\hline
\end{tabular}
\end{center}
\tablebib{
(1) \citet{zijlstra89}; 
(2) \citet{sevenster97a};
(3) \citet{vandsteene01}; (4) \citet{tafoya09};
(5) \citet{gomez09}; (6) this paper
}
\\

\tablefoottext{a}{Designation of galactic planetary nebulae in the catalog of \citet{acker92}}
\tablefoottext{b}{The position uncertainty estimated for radio
  interferometric observations is of the order of 1/10 of the
  synthesized beam, to be conservative we considered a positional
  error of $\sim$1/5 of the synthesized beam (See Sect. \ref{association}).}
\tablefoottext{c}{Literature reference for the quoted positions.}
\tablefoottext{d}{Simultaneous OH and continuum observations}
\tablefoottext{e}{The relationship of JaSt 23 with IRAS 17371$-$2747 is uncertain.}

\end{table}
\end{landscape}

\begin{landscape}
\begin{table}
\begin{center}
\caption{Possible OHPNe \label{possiblesources}}
\begin{tabular}{lcllccrccc}
\hline \hline 
IRAS & PN G   &
\multicolumn{2}{c}{Continuum position} & Error  &
\multicolumn{2}{c}{OH position}  & Error & Position
difference & References \\
& &   RA (J2000) & Dec (J2000)  & ($''$) & RA (J2000) & Dec (J2000)  &   ($''$) &($''$) &        \\
\hline
 17150$-$3754 &  &  17 18 28.97 & $-$37 58 01.9 & 0.4 & 17 18 28.922& $-$37 58 01.79 & 2.0 & 0.6 & 3 \\
 17168$-$3736 &  & 17 20 15.08 & $-$37 39 31.5 & 1.1 & 17 20 15.047 & $-$37 39 34.31 & 2.4 & 2.8 & 2,3 \\
 17221$-$3038 & 356.1+02.7 &   17 25 19.37 & $-$30 40 42.3 & 0.6 & \tablefootmark{a}  &  & & & 3 \\
 17375$-$2759 && 17 40 38.57 & $-$28 01 00.4 & 0.7  & 17 40 38.587 & $-$28 01 01.08 & 2.4 & 0.7 & 1, 2 \\   
\hline
\end{tabular}
\end{center}
\tablebib{
 (1) \citet{pottasch88};
 (2) \citet{sevenster97a};
 (3) this paper
}
 \\
 \tablefoottext{a}{No interferometric OH observation available}

\end{table}
\end{landscape}

\begin{table}
\begin{center}
\caption{Misclassified OHPNe \label{misclassified}}
\begin{tabular}{lcccc}
\hline \hline 
IRAS  &  \multicolumn{2}{c}{OH position}  & References & Notes\\
      & RA (J2000) & Dec (J2000)   &  &      \\
\hline
17207$-$2856 &     &  &  & \tablefootmark{a}  \\
17375$-$3000 & 17 40 43.339 & $-$30 02 04.86  & 1  &   \\
17418$-$2713 & 17 44 58.723 & $-$27 14 43.37  & 1 & \tablefootmark{b}  \\
17443$-$2949 & 17 47 35.442 & $-$29 50 53.32  & 2 &  \tablefootmark{b}  \\
17580$-$3111 & 18 01 20.399 & $-$31 11 20.55   & 2  & \tablefootmark{c}  \\

\hline
\end{tabular}
\end{center}
\tablebib{
(1) \citet{sevenster97a};
 (2) \citet{gomez08} 
}
\tablefoottext{a}{Neither radio continuum nor OH emission was detected}
\tablefoottext{b}{Possible AGB star}
\tablefoottext{c}{Possible post-AGB star}
\end{table}

\end{document}